\documentclass[a4paper,pra,twocolumn,superscriptaddress]{revtex4}

\usepackage[english]{babel}
\usepackage[utf8]{inputenc}
\usepackage{graphicx,epsfig,xcolor}
\usepackage{latexsym}
\usepackage{amsfonts}
\usepackage{amsmath}
\usepackage{textcomp}
\usepackage{hyperref}
\usepackage{placeins}
\def\<{\langle}
\def\>{\rangle}
\def\{{\lbrace}
\def\}{\rbrace}
\def\({\left(}
\def\){\right)}
\def\beq{\begin{equation}}
\def\eeq{\end{equation}}
\def\c#1{\mathcal{#1}}
\def\eff{\text{eff}}

\def\bx{\textbf{x}}

\begin{document}

\title{First-Passage Percolation under extreme disorder: from
  bond-percolation to Kardar-Parisi-Zhang universality}

\author{Daniel Villarrubia}
\affiliation{Dpto. F\'{\i}sica Matem\'atica y de Fluidos, UNED, Spain}

\author{Iván Álvarez Domenech}
\affiliation{Dpto. F\'{\i}sica Matem\'atica y de Fluidos, UNED, Spain}

\author{Silvia N. Santalla}
\affiliation{Dpto. F\'{\i}sica \& GISC, Universidad Carlos III de Madrid, Spain}

\author{Javier Rodr\'{\i}guez-Laguna}
\affiliation{Dpto. F\'{\i}sica Fundamental, UNED, Spain}

\author{Pedro C\'ordoba-Torres}
\affiliation{Dpto. F\'{\i}sica Matem\'atica y de Fluidos, UNED, Spain}

\begin{abstract}

We consider the statistical properties of arrival times and balls on
first-passage percolation (FPP) $2D$ square lattices with strong
disorder in the link-times. A previous work showed a crossover in the
weak disorder regime, between Gaussian and Kardar-Parisi-Zhang (KPZ)
universality, with the crossover length decreasing as the noise
amplitude grows. On the other hand, this work presents a very
different behavior in the strong-disorder regime. A new crossover
length appears below which the model is described by bond-percolation
universality class. This characteristic length scale grows with the
noise amplitude and diverges at the infinite-disorder limit. We
provide a thorough characterization of the bond-percolation phase,
reproducing its associated critical exponents through a careful
scaling analysis of the balls. This is carried out through to a
continuous mapping of the FPP passage time into the occupation
probability of the bond-percolation problem. Moreover, the crossover
length can be explained merely in terms of properties of the link-time
distribution. The interplay between the new characteristic length and
the correlation length intrinsic to bond-percolation determines the
crossover between the initial percolation-like growth and
theasymptotic KPZ scaling.
\end{abstract}

\date{December 15, 2019}

\maketitle

\section{Introduction}
\label{sec:intro}

Geometry on random manifolds presents both applied and fundamental
interest, with applications ranging from the physics of polymers and
membranes \cite{Nelson,Boal} to quantum gravity
\cite{Itzykson_Drouffe,Booss.book,Ambjorn_97}. Specifically, {\em
  geodesics} and {\em isochrones} on random manifolds, corresponding
to non-Euclidean versions of straight lines and circumferences,
present a very rich behavior \cite{Santalla_15,Santalla_17}. It was
recently shown that, in the case of random surfaces which are flat in
average and with short-range correlations in the curvature, geodesics
present fractal structure, governed by exponents corresponding to the
celebrated Kardar-Parisi-Zhang (KPZ) universality class
\cite{Kardar_86} describing random interfacial growth
\cite{Barabasi,Krug_97,Krug_10,Halpin_15}. Specifically, the lateral
deviation of a geodesic joining two points separated an Euclidean
distance $L$ scales like $L^{1/z}$, where $z=3/2$ is the dynamical
exponent of KPZ. Moreover, the deviation of the arrival times scales
like $L^\beta$, with $\beta=1/3$, and their fluctuations follow the
Tracy-Widom distribution for the lowest eigenvalues of random unitary
matrices \cite{Praehofer_02,Takeuchi_11,Corwin_13}.

When the manifold is discretized the problem is called {\em
  first-passage percolation} (FPP)
\cite{Hammersley_65,Howard_04,Auffinger_17}. Given an undirected
lattice, e.g. $\mathbb{Z}^2$, a randomly chosen link-time $t$ is
assigned to each edge between neighboring nodes. Link-times are
independent and identically distributed (i.i.d.) random variables with
common probability density function $f(t)$ and cumulative distribution
function $F(t)$. Notice that for the model to present the structure of
a metric space we must assume $F(0)=0$.  This problem bears a strong
relation with the {\em directed polymers in random media} (DPRM)
\cite{Kardar_87,Krug_91,Halpin_95}, where we are asked to find the
minimal energy configuration for a polymer of fixed length on a random
surface. The most relevant difference between them is the fixed length
constraint, which does not hold for geodesics. FPP results have been
successfully applied to magnetism \cite{Abraham_95}, wireless
communications \cite{Beyme_14}, ecological competition
\cite{Kordzhakia_05} and molecular biology \cite{Bundschuh_00}.

The main objects of study in FPP are geodesics, i.e. minimal time
paths joining pairs of points, and balls $B(T)$ given by the set of
nodes which can be reached from the origin in a time less than
$T$. There are some rigorously proved results, such as the {\em
  Galilean invariance}, $z(1+\beta)=2$ \cite{Chatterjee_13}. It has
also been proven (\emph{shape theorem}) that, when the lattice
structure is properly smoothed out, the ball $B(T)$ grows linearly
with $T$ and has an asymptotic shape $B_0$ which is
non-random. Moreover, for suitable conditions on the moments of $F$,
$B_0$ is a convex set with nonempty interior and it is either compact
or equals all of $\mathbb{R}^d$ (see \cite{Kesten_87} and references
therein).

FPP presents a whole new set of features over the continuous version
of the problem, such as a direction-dependent crossover associated
with the so-called {\em geodesic degeneracy} \cite{Cordoba_18}, i.e.
the number and structure of the geodesics joining two points in
absence of noise. A characteristic length $d_c$ was found both in the
square lattice and in random Delaunay triangulations, determining a
crossover between Gaussian and KPZ behavior. This crossover length
decreases as the noise amplitude increases following $d_c\sim ({\rm
  CV})^{-2}$, where CV is the {\em coefficient of variation} of the
link-time distribution, defined as ${\rm CV}\equiv s/\tau$, where
$\tau$ and $s$ are respectively the mean value and deviation of the
link-times \cite{Cordoba_18}. Those results are only valid when
$d_c>1$, i.e. for $\tau>s$ or, in other terms, in the {\em
  weak-disorder regime}.  For example, all uniform noise distributions
fall into this regime.

Yet, for a strong noise, i.e. $s \gg\tau$, the characteristic length
$d_c<1$ can not play the same role. In some relevant cases, $s$ and
$\tau$ might even be ill-defined. Indeed, in the {\em strong-disorder
  regime}, link-times encompass several orders of magnitude. A limit
case is given by the Bernoulli distribution, in which link-times can
be zero with probability $p=F(0)$. If we identify now zero-time links
with open bonds in a connected lattice we are effectively mapping our
system to a problem of {\em bond-percolation}
\cite{Kesten_87,Halpin_95,Stauffer_03}. Thus, {\em critical FPP} is
defined by condition $F(0)=p_c$ \cite{Damron_17, Damron_19, Yao_18,
  Kesten_97, Yao_14} and \emph{supercritical FPP} is defined by
$F(0)>p_c$ \cite{Zhang_95, Garet_07,Yao_13}, where $p_c$ is the
critical probability in bond-percolation \cite{Stauffer_03}. For
$F(0)\geq p_c$ there exists an infinite connected set of edges with
zero crossing-time so that traveling across this infinite cluster
costs no time \cite{Aizenman_87}. As a consequence, the \emph{route}
between two lattice nodes will always stay within the infinite cluster
except for a few edges \cite{Zhang_95}. Furthermore, it has been shown
that the asymptotic shape $B_0$ equals all of $\mathbb{R}^d$ if and
only if $F(0)\geq p_c$ \cite{Kesten_86, Chayes_86}.

Much effort in critical FPP has been devoted to the analytical study
of the \emph{time constant} $\mu$, defined as the limit of the passage
time to a site (or to some boundary) normalized by its distance from
the origin, as the distance goes to infinity. This limit has been
proven to exist for suitable conditions on the moments of $F$
\cite{Smythe_78}, and $\mu=0$ if and only if $F(0)\geq p_c$
\cite{Kesten_87}. Indeed, the existence of large clusters with
zero-weight edges leads to a passage time that grows at most
logarithmically with distance, yielding a zero time constant
\cite{Damron_19,Yao_18,Damron_17}. The work has focused on the
characterization of the asymptotics of this arrival time by conditions
on the distribution function $F$ \cite{Damron_17,
  Damron_19,Yao_18,Kesten_97,Yao_14}, yielding relevant results such
as a universal expression for the time constant on $2D$ lattices
\cite{Damron_19}, the proposition of a central limit theorem for the
passage time \cite{Yao_18, Kesten_97} or a relation between critical
FPP and invasion percolation \cite{Damron_17}.

The geometry of the minimal paths has also been a matter of study \cite{Giles_19}. In the early 80's, Ritzenberg and Cohen considered shortest paths on a {\em percolation   cluster} beyond criticality, i.e. spanning an infinite number of
nodes \cite{Ritzenberg_84}, computing their fractal
dimension. Kerstein and coworkers considered FPP with two possible
link-times, a slow one and a (extremely) fast one
\cite{Kerstein_85,Kerstein_85b,Kerstein_86}. As the density of slow
links increases, they describe a crossover from {\em chemical} to {\em
  contact propagation}. Chemical propagation describes geodesics that
only employ fast links, while contact propagation refers to the use of
slow links to jump from cluster to cluster. In the first case, the
conduction rate is limited by the geodesic {\em tortuosity}, i.e. the
average number of links along a typical fast path. For contact
propagation, on the other hand, the conduction rate is limited by the
ratio of slow bonds. The crossover between both regimes takes place at
the percolation threshold. Geodesics on critical and super-critical
percolation clusters were studied by several authors
\cite{Zhang_95,Garet_07,Yao_13,Damron_17,Yao_18,Garet_10}, and a
correspondence between geodesics on critical percolation clusters and
Schramm-Loewner evolution curves has been recently put forward
\cite{Pose_14}, lending support to the idea that they might be
conformally invariant \cite{Di_Francesco}.

In this article, we characterize the statistical properties of arrival
times and balls on strongly disordered networks. Our purpose is to
study the behavior of the model close to the critical case discussed
so far, but keeping the condition $F(0)=0$ necessary for it to
represent a metric space. We propose a mapping of the passage time
into the probability $p$ of a bond being open, that allows us to map
the FPP problem into a family of bond-percolation problems. Making use
of detailed numerical simulations, we show how the scaling exponents
of percolation theory map into those of the geodesic behavior below a
certain crossover length, above which the geodesics attain the
standard KPZ behavior. This new crossover length increases with the
noise amplitude and seems to diverge for infinite noise, which leads
us to conjecture that this limit might be related to the critical or
supercritical FPP cases, depending on the link-time distribution.

This article is organized as follows. Section \ref{sec:model} presents
the first-passage percolation model and our basic assumptions. Next,
in section \ref{sec:passage_time_fluctuations} we compare the
statistical properties of the arrival times in the weak and strong
disorder regimes. The mapping of the FPP problem into bond percolation
is addressed in section \ref{sec:percolation}, where we also present a
comprehensive scaling analysis of the FPP balls that recovers the
critical exponents of percolation. In section \ref{sec:chain_model} we
propose a model for the characteristic length that controls the extent
of the percolation domain. Large scale behavior and the crossover
towards standard KPZ scaling is described in Sec. \ref{sec:crossover},
in which we also discuss the transition between the weak and strong
disorder regimes. Section \ref{sec:conclusions} is devoted to a
summary of our conclusions and our ideas regarding future work.


\section{Model and definitions}
\label{sec:model}

Let us consider a $L\times L$ square lattice (odd $L$), with nodes
$\bx_i$ and a central node $\bx_0=(0,0)$. We assign a link-time
$t(\bx_i,\bx_j)$ to each pair of nearest-neighbor nodes, $\bx_i$ and
$\bx_j$. Given a path $\Gamma=\{\bx_0,\bx_1,\cdots,\bx_m\}$ joining
the center to site $\bx_m$, in order to
traverse that path we would need a time

\beq
T_{\Gamma}(\bx_m)=\sum_{i=1}^{m} t(\bx_{i-1},\bx_{i}).
\label{eq:arrival_time}
\eeq
We define the {\em arrival time} as the minimal value over all
paths reaching arbitrary node $\bx$ from $\bx_0$:

\beq
T(\bx)\equiv \min_\Gamma\{T_{\Gamma}(\bx)\},
\label{eq:passage_time}
\eeq
and the corresponding path is the \emph{geodesic} or \emph{optimal
  path} to that point. For a generic distribution of link-times the
geodesic will be unique, but geodesic degeneracy can play an important
role in some cases \cite{Cordoba_18}. Let us remark that arrival times
can be efficiently obtained for all sites making use of {\em
  Dijkstra's algorithm}. Finally, we also define the ball $B(T)$ as
the set of nodes which can be reached in a time smaller than a certain
value $T$.

We consider the case in which link-times are independent and
identically distributed (i.i.d.) random variables with common
probability density function $f(t)$ and cumulative distribution function $F(t)$, with the only constraint of positivity: $F(0)=0$. If they exist, the link-time distribution is characterized by a mean value $\tau$ and a variance $s^2$,

\medskip

As discussed in the introduction, it has been shown \cite{Cordoba_18}
that for weakly disordered distributions with $\tau > s$ (CV$<1$)
there exists a crossover length given by $d_c=a(\tau/s)^2>1$, with the
scale factor $a=O(1)$ depending on the geometrical properties of the
lattice (e.g. $a\simeq1/3$ for the square lattice), such that arrival
times to sites along the axis at a distance $x\ll d_c$ follow Gaussian
statistics, and KPZ statistics for $x\gg d_c$. Indeed, in that case
the geodesic will not deviate from the straight line up to a distance
$\sim d_c$. This behavior is expected to occur along the
non-degenerate directions of regular lattices and also in low
degeneration random lattices such as Delaunay triangulations
\cite{Cordoba_18}. On the other hand, for disordered distributions
with $\tau <s$ (CV$>1$) thus giving $d_c<1$ we enter the {\em strong
  disorder} regime, and the statistical properties of the arrival
times change considerably.

Uniform link-time distributions have necessarily $d_c> 1$ (note that
CV$<3^{-1/2}$), but other distributions interpolate smoothly between
these two regimes. That is the case of the three probability
distributions employed in this work: Weibull (Wei$(\lambda,k)$),
Pareto (Par$(t_m,\alpha)$) and Log-Normal (LogN$(\mu,\sigma)$), whose
definition and basic properties are listed in Appendix
\ref{appendix:distrib}. The three distributions are characterized by
a {\em shape} parameter determining the amplitude of the noise: the
power-law exponents $k$ and $\alpha$ in the Weibull and Pareto
distributions respectively, and the variance $\sigma$ in the
Log-Normal. As we show there, in all cases $d_c$ is a monotonic
function of it, increasing for the first two cases and decreasing for
the Log-Normal.

In order to unify our description we introduce the concept of
\emph{order parameter} to identify the distribution parameter which,
when decreasing, makes the strength of disorder increase (by
convention) in a monotonic way. We thus identity $k$, $\alpha$ and
$\sigma^{-1}$ as the corresponding order parameters of the above
distributions. For the sake of clarity we will denote this order
parameter by $o$. As a result of this convention $d_c$ is an
increasing monotonic function of $o$. We also introduce the symbol
$o^{\star}$ to denote the value of the order parameter yielding
$d_c=1$: $d_c(o^{\star})=1$. In other words, $o^{\star}$ determines
the crossover point between the weak ($o>o^{\star}$ and $d_c>1$) and
the strong ($o<o^{\star}$ and $d_c<1$) disorder regimes.

There is an interesting limit of the model which will be considered as a reference, the so-called \emph{homogeneous case} in which link-times have uniform value $\tau_0$, and which corresponds to the delta distribution $f(t)=\delta(t-\tau_0)$. As we show in Appendix \ref{appendix:distrib}, the homogeneous case is obtained at the limit $o\rightarrow \infty$ (i.e. ``infinite'' order) of the above distributions (Eq. \eqref{eq:homogeneous_case}).

\begin{figure*}
  \hbox to \hsize{\hfill
  \includegraphics[width=\columnwidth]{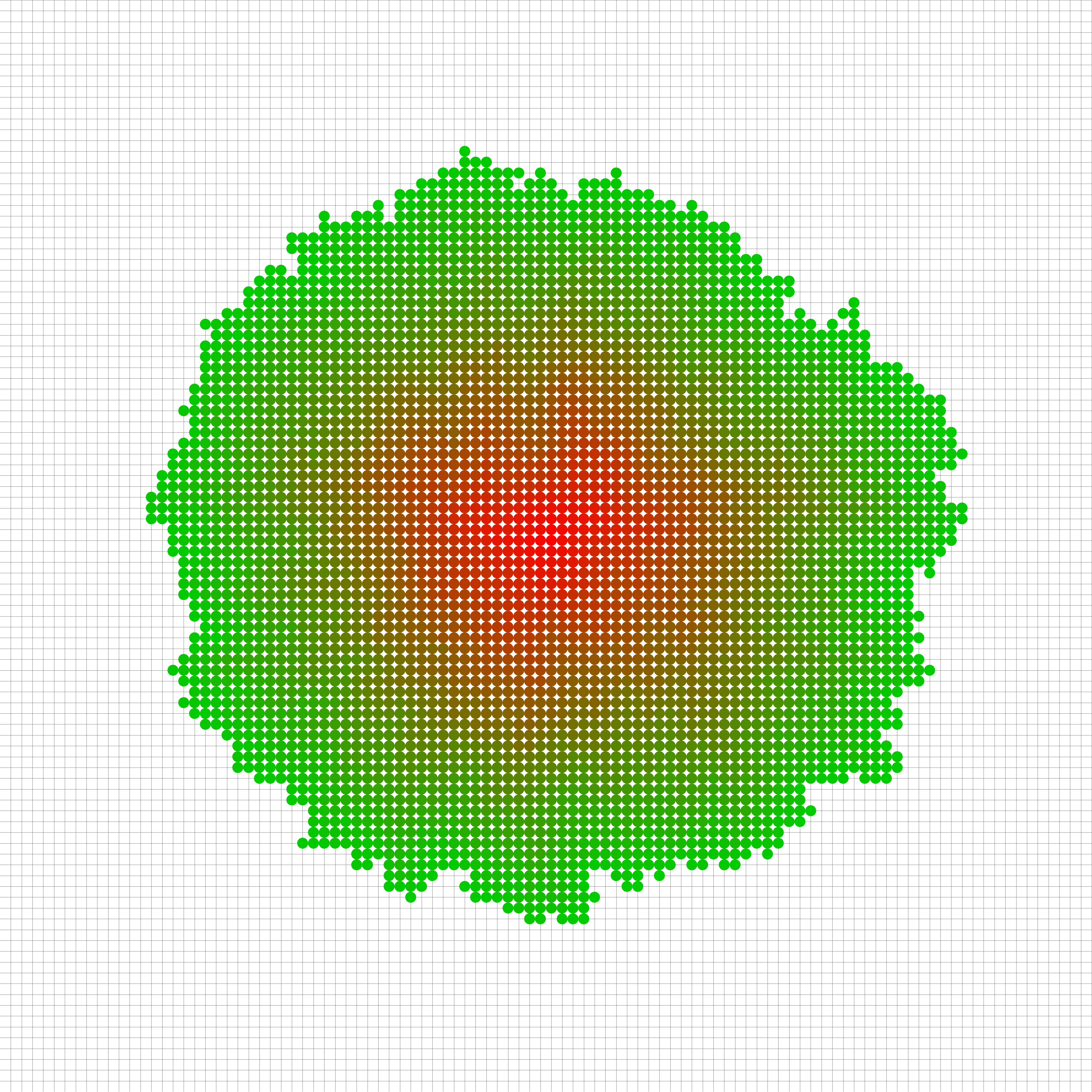}\hfill
  \includegraphics[width=\columnwidth]{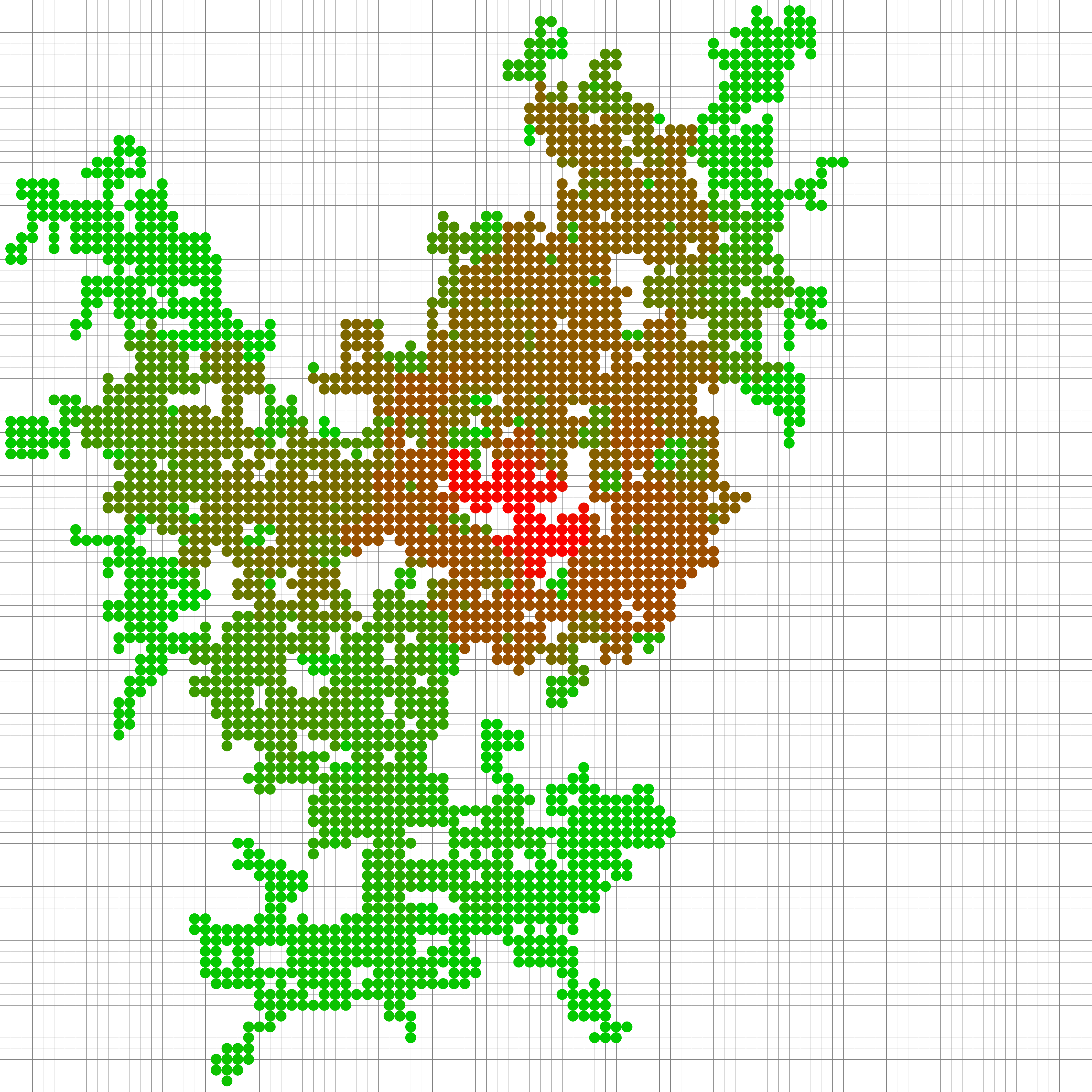}\hfill}
  \caption{(Left) FPP ball corresponding to $T=25$ on a square
    lattice, for a link-time Weibull distribution Wei$(1,2)$. Colors
    are associated to different arrival times. (Right) FPP ball
    corresponding to $T=0.1$ when the link-times are drawn from
    Wei$(1,0.1)$. Notice that the crossover value of the order
    parameter is $k^{\star}\approx 1.79$. }
  \label{fig:illust}
\end{figure*}

\medskip

Before going in-depth into the analysis of the model, it is worth having a look at the different behaviors at both sides of this divide.
The left panel of Fig. \ref{fig:illust} shows the FPP ball $B(T)$ obtained for $T=25$ on a $L=101$ lattice whose
link-crossing times are drawn from a Weibull distribution with $k=2$, for which $d_c\approx 1.22$. The final ball appears {\em rough}, and indeed this roughness can be shown to correspond to the well-known KPZ
universality class. Colors provide information about the arrival time to all sites within the ball, which we can see
to behave in a reasonable smooth way. On the other hand, the right panel corresponds to $T=0.1$  for a distribution with $k=0.1$, for which $d_c\approx 2\cdot 10^{-6}$. Notice that the
aspect of the ball is much less round, with abundance of {\em
  cavities}. Another salient feature is the {\em sharpness} of the
color gradation. Instead of a smooth variation, we can determine clear
boundaries within the ball.


\section{Fluctuations of the arrival times: from weak to strong disorder}
\label{sec:passage_time_fluctuations}

Let us explore numerically the statistical properties of the arrival times using a $L\times L=2001\times 2001$ lattice and link-times drawn from the distributions presented above with different values of the order parameter,
in order to interpolate smoothly from the weak to the strong-disorder regimes. Moreover, we will always average over $N=10^4$ samples,  unless otherwise stated. As in Fig. \ref{fig:illust}, most of the results displayed hereafter will be obtained from the Weibull distribution, for which the crossover value of the order parameter is $o^{\star}=k^{\star}=1.79$ (Appendix \ref{appendix:distrib}). Unless otherwise stated, it must be assumed that they hold for the other distributions. If results are distribution dependent we will address each distribution separately.

Let us consider the average arrival time in units of the average
link-time, $\<T\>/\tau$, for sites on the axis as a function of the
distance to the center, $x=\|\bx\|$.  In the weak disorder regime
\cite{Cordoba_18} this magnitude grows linearly with the distance for
$x\gg d_c$, with a slope that continuously decreases as the order
parameter decreases. This slope represents the inverse of the
normalized velocity of growth, closely related to the \emph{time
  constant} $\mu$ discussed in the introduction. It is bounded from
above by the trivial value obtained in the homogeneous case
($o\rightarrow \infty$): $1$ in the axis and $\sqrt{2}$ for the
diagonal, leading to balls with diamond shapes. As the order parameter
decreases the slopes found in all lattice directions decrease
(velocity increases), and when $o$ approaches the crossover value
$o^{\star}$ ($d_c\rightarrow 1$), they become equal, thus explaining
the circular shape of the balls in average (see e.g. the ball shown in
left panel of Fig. \ref{fig:illust}).

\begin{figure}
  \includegraphics[width=\columnwidth]{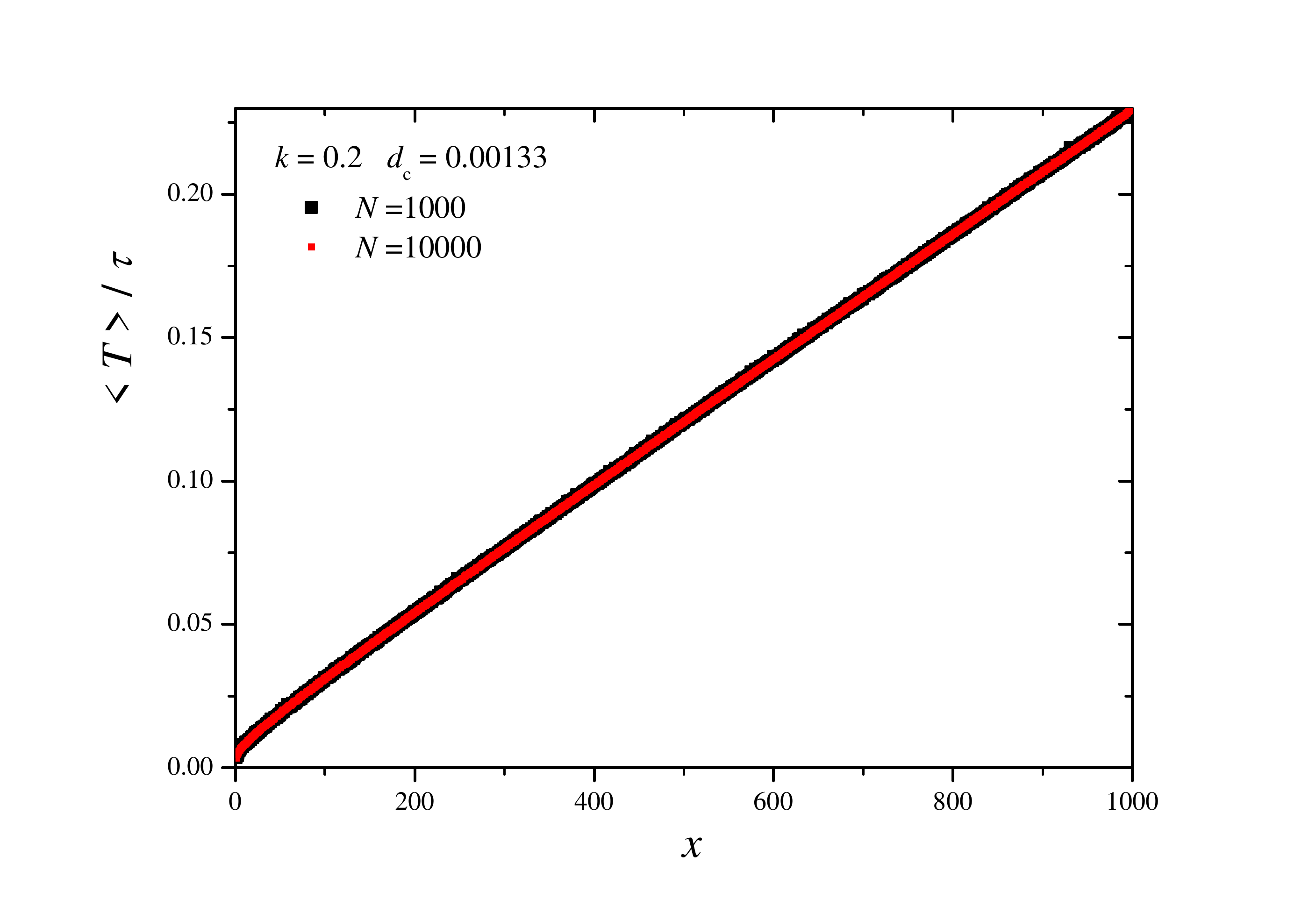}
  \includegraphics[width=\columnwidth]{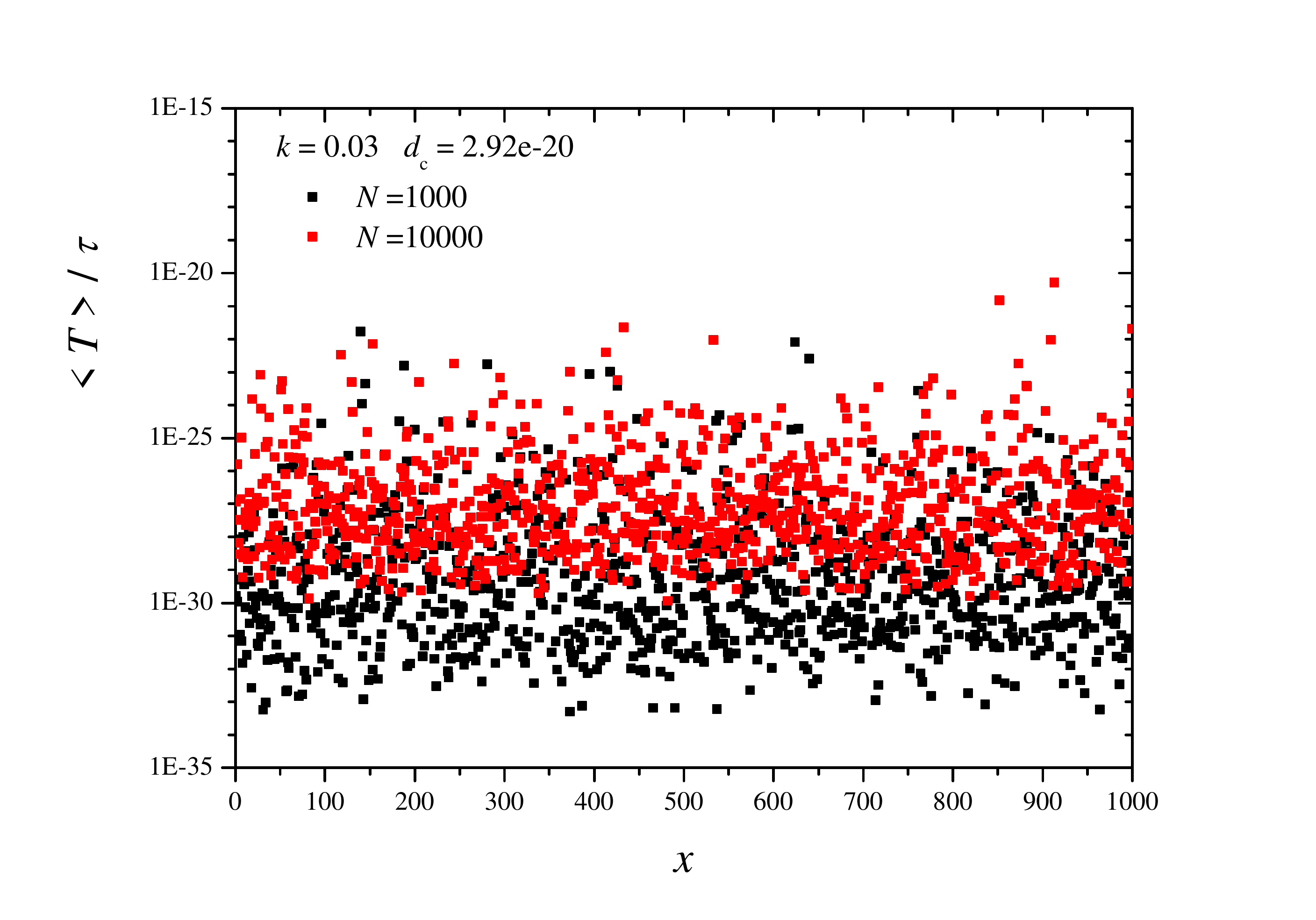}
  \caption{Average passage time, rescaled to the mean link-time
    $\tau$, to nodes at distance $x$ along the axis for the Weibull
    distribution with $k=0.2, $ (top) and $k=0.03$ (bottom). Results
    from two ensembles $N=1000$ and $N=10000$ have been represented in
    both cases with black and red symbols respectively.}
  \label{fig:Weibull_time_axis}
\end{figure}

The top panel of Fig. \ref{fig:Weibull_time_axis} shows $\<T\>/\tau$
when the link-times follow the Weibull distribution for
$k=0.2$. The arrival time seems to approach a linear growth with distance giving
a slope much smaller than one. For reasons that will become apparent soon, we have
employed two different sample sizes: $N=10^3$ (black symbols) and
$N=10^4$ (red symbols), which coincide perfectly. The lower panel of
Fig. \ref{fig:Weibull_time_axis} shows the values of $\<T\>/\tau$ as a
function of the distance when the link-times are drawn from a much
lower value of the order parameter $k=0.03$, i.e. deep in the
strong-disorder regime ($k\ll k^{\star}$). In this case, the times of
arrival appear scattered, without a clear dependence on the
distance. Moreover, the data for $N=10^3$ (black) and $N=10^4$ (red)
are statistically different. This fact suggests that arrival times in
this regime are very difficult to sample, pointing to an extremely
broad distribution.

\begin{figure}
  \includegraphics[width=\columnwidth]{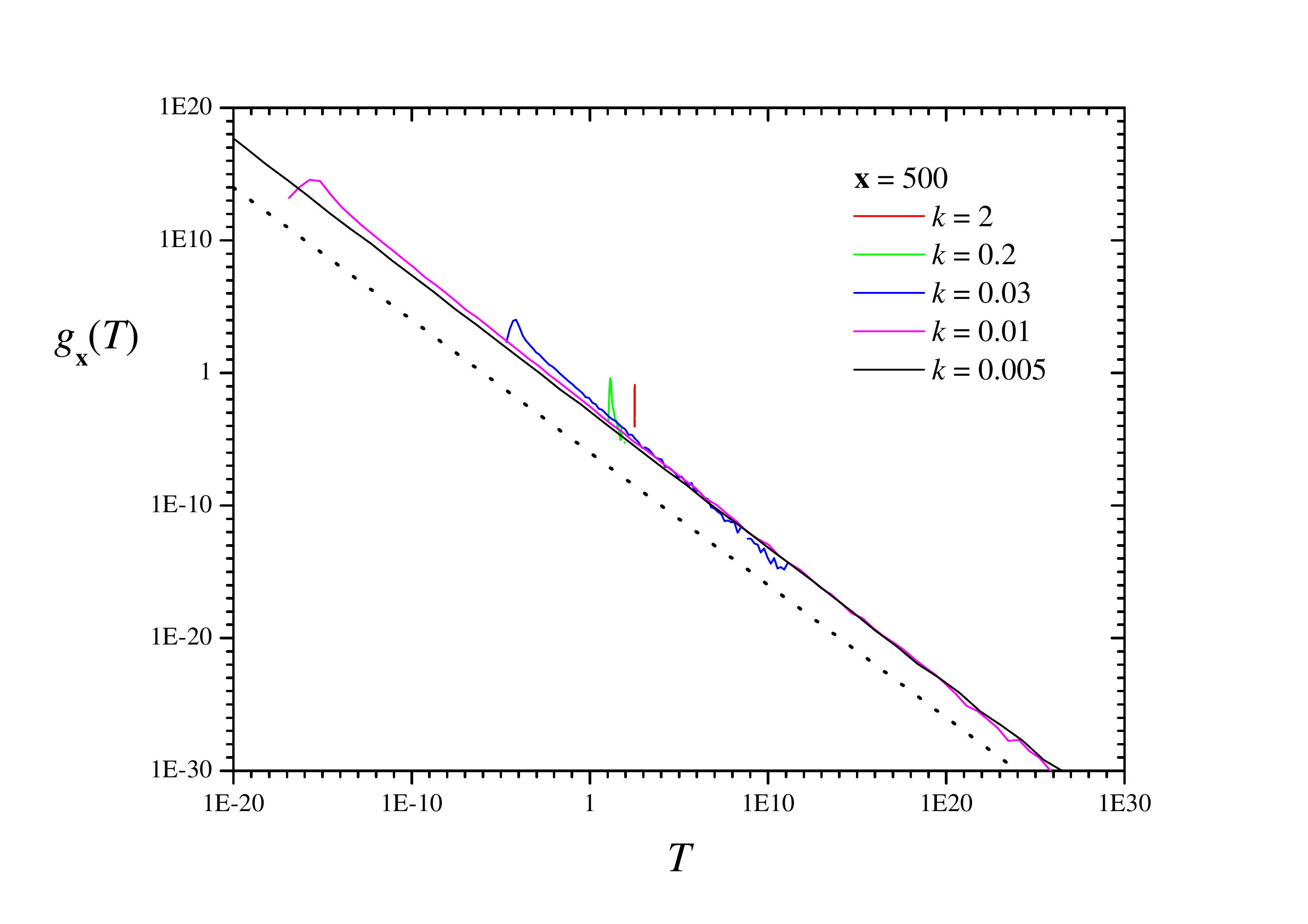}
  \includegraphics[width=\columnwidth]{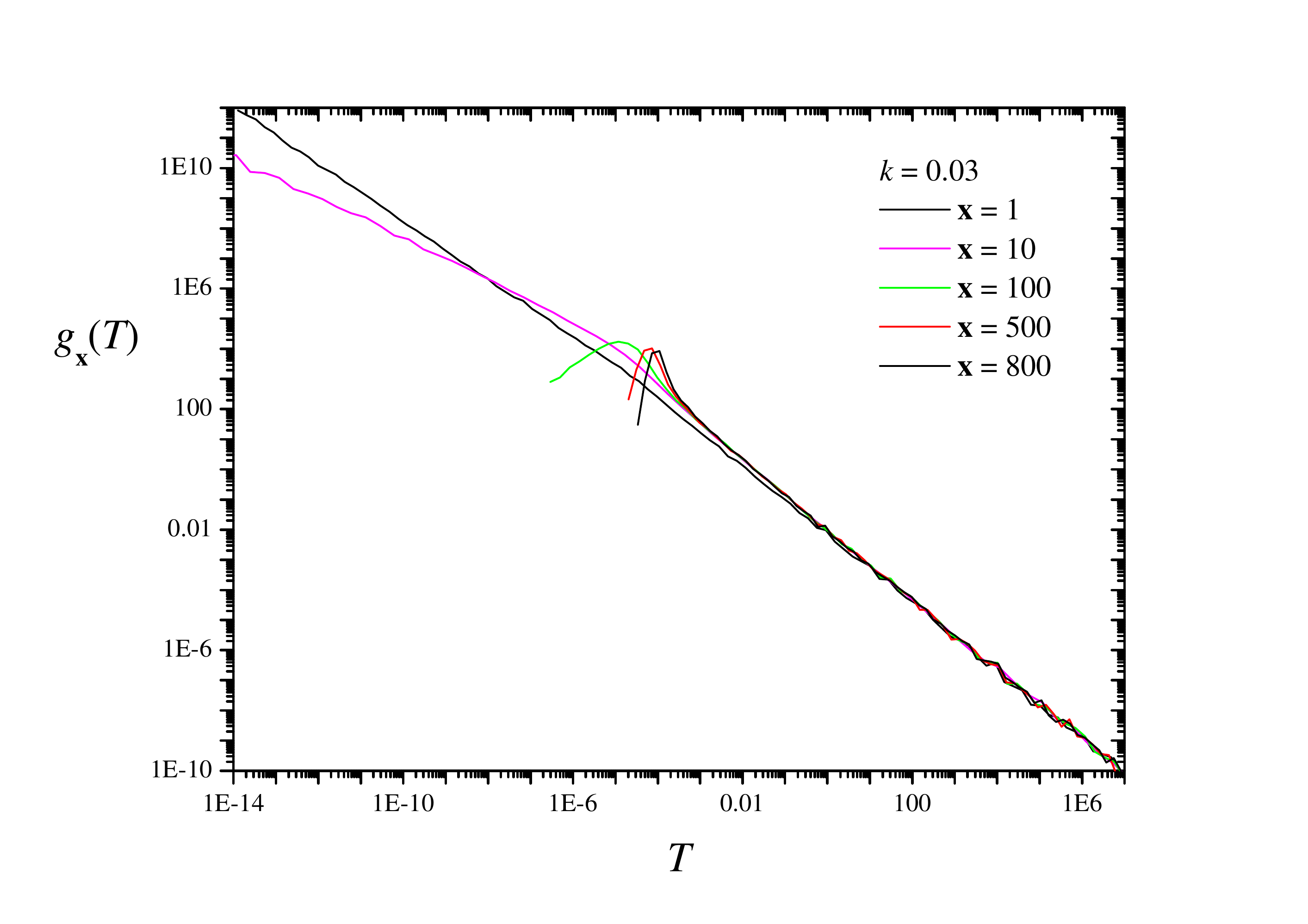}
  \caption{Histogram of passage time $T$ to: (top) site at distance
    $x=500$ in the axis for different values of $k$; (bottom)
    different sites along the axis for same $k=0.03$. Dotted line in top panel indicates the scaling $\sim T^{-1}$}
  \label{fig:histograma_T_Weibull}
\end{figure}

To understand this behavior we have displayed in
Fig. \ref{fig:histograma_T_Weibull} the distribution of the arrival
times to a given node $\bx$, denoted by $g_{\bx}(T)$. In the top panel
we have considered a fixed position along the axis, $x=500$ and
link-time distributions of the form Wei(1,$k$) for different values of
$k$. The bottom part of Fig. \ref{fig:histograma_T_Weibull} displays
the results for the strongly disordered case with $k=0.03$ at
different points along the axis.

In the weak-disorder regime, the distribution of arrival times at
distance $x\gg d_c$ has been shown to follow the Tracy-Widom
distribution for the Gaussian unitary ensemble (GUE)
\cite{Praehofer_02,Takeuchi_11,Corwin_13,Cordoba_18}, which appears in
the top panel of Fig. \ref{fig:histograma_T_Weibull} (case $k=2$) in
the form of a small and sharp peak (notice the log scale). However, as
the disorder strength increases the distribution becomes right-skewed
with the arrival time spanning an increasing range of orders of
magnitude. There is still a well defined mode that moves towards
smaller values and which is followed by a increasingly longer tail
displaying a power-law scaling $\sim T^{-1}$ as $k\rightarrow 0$. As as consequence, for extreme disorder the mean
arrival time is not well-defined and its average value is dominated by
the largest value: $\<T\>(\bx)={1\over N} \sum_{i}^{N} T_i(\bx)\simeq
{1\over N} T_{\max N}(\bx)$, where $T_{\max
  N}(\bx)=\max_i\{T_i(\bx)\}$, which certainly depends on the sample
size thereby explaining the results displayed in
Fig. \ref{fig:Weibull_time_axis} (bottom).

Let us now focus on the dependence of the passage-time distribution on
distance for highly disperse distributions (case $k=0.03$, bottom
panel of Fig. \ref{fig:histograma_T_Weibull}). We observe that the
right tails of the skewed distributions merge into a single decay
regardless of the position of the target point. This result, together
with the max principle stated above, explain the independence of the
average time $\<T\>$ with distance displayed in
Fig. \ref{fig:Weibull_time_axis} (bottom). Furthermore, as we move
away from the center node, the distributions seem to approach a limit
function which reveals that the minimum arrival time required to reach
any lattice site is independent of its position. The existence of such
minimal arrival time is an evidence of criticality in the model.

\medskip

Another physical quantity of immediate interest in the study of the
FPP model is the standard deviation of the arrival times. In the KPZ
regime ($o>o^{\star}$ and for $x\gg d_c$) it scales as $\sigma_T \sim x^\beta$,
with $\beta=1/3$, following the Family-Vicsek Ansatz. This standard deviation also corresponds to the roughness of the balls
\cite{Santalla_15} and is analogous to the free energy
fluctuations in directed polymers in random media (DPRM)
\cite{Halpin_95}.

\begin{figure}
  \includegraphics[width=9.5cm]{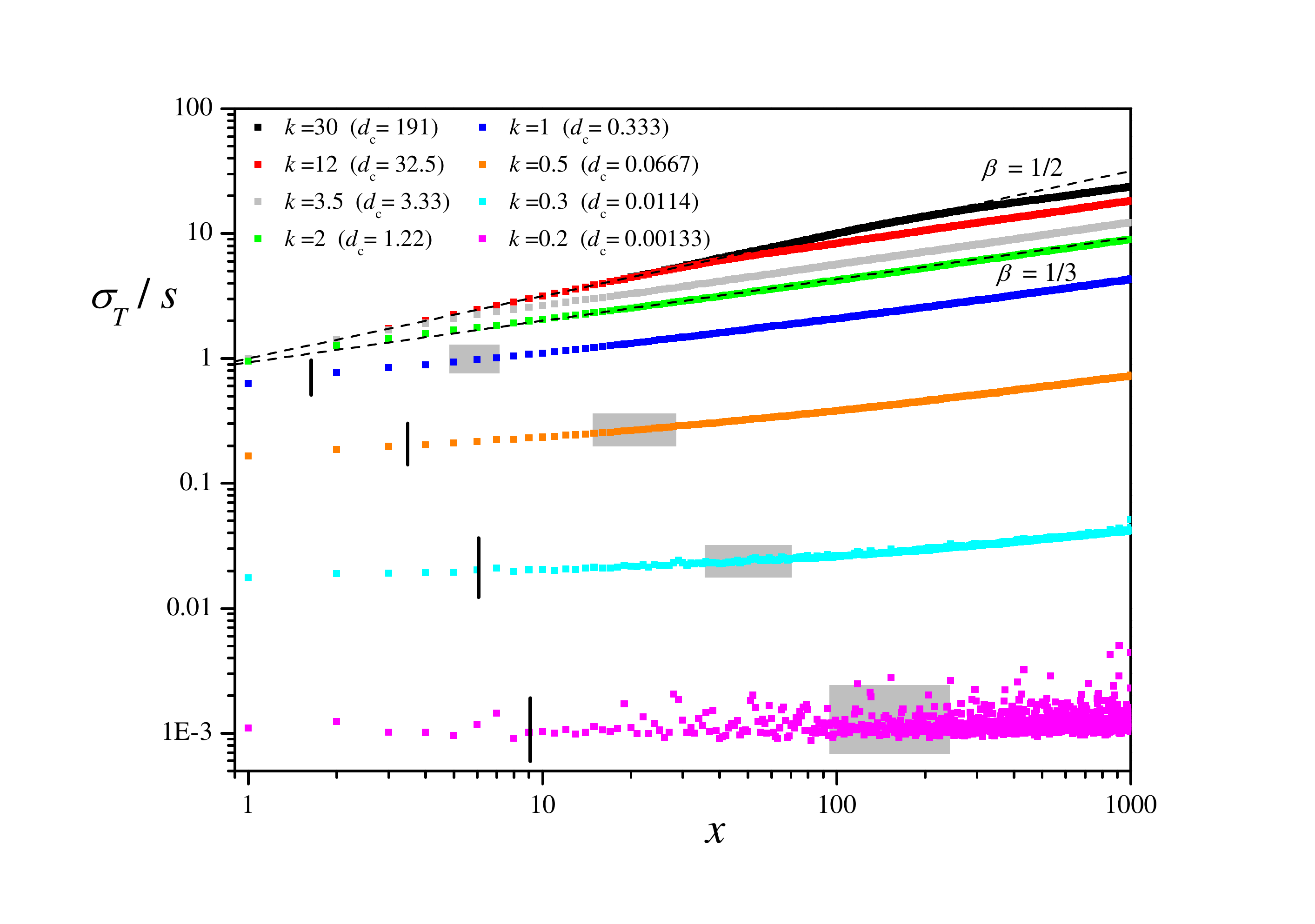}
  \caption{Standard deviation $\sigma_T$ of the arrival time to nodes
    at distance $x$ in the axis of the square lattice with link-times
    following a Wei(1,$k$) distribution, normalized to the standard
    deviation of the link-time distribution $s$. The corresponding
    values of $d_c$ are calculated from Eq. \eqref{eq:wei_dc}. Broken
    lines represent Gaussian pre-asymptotic growth ($\beta=1/2$) and
    KPZ scaling ($\beta=1/3$). Vertical segments in curves with
    $d_c<1$ ($k<k^{\star}$) indicate the value of length $\ell_o(T_c)$
    obtained from the chain model presented in
    Sec. \ref{sec:chain_model}. Grey rectangles represent the interval
    $[\ell_o(T(p=0.99)), \ell_o(T(p=0.999))]$, as discussed in
    Sec. \ref{sec:crossover}.}
  \label{fig:passage_time_deviation_axis_Weibull}
\end{figure}

In Fig. \ref{fig:passage_time_deviation_axis_Weibull} we have plotted
the standard deviation of the arrival time $\sigma_T$, in units of the
deviation of the link-time distribution, $s$, to nodes in the axis as a function of their distance $x$ to the center. In all cases, we are employing Weibull distributions Wei(1,$k$), for a range of values of $k$, which are shown along with the corresponding $d_c$. For $k>k^{\star}$ ($k=2$ up to $k=30$) we observe the well-known crossover from Gaussian towards KPZ scaling, with a behavior $\sigma_T\sim x^{1/2}$ for $d\ll
d_c$ and $\sigma_T\sim x^{1/3}$ for $d\gg d_c$
\cite{Cordoba_18}. Moreover, for $x=1$ we have $\sigma_T/s\approx 1$, because
the geodesic from a site to its neighbor usually consists of traversing the
link joining them. On the other hand, we observe that for $k<k^{\star}$, $\sigma_T/s$ at $x=1$ is always below 1, and decreases very fast as $k\to 0$,
implying that the geodesic between a site and its neighbor may be
non-trivial. This fact suggests that geodesics in the strong-disorder
regime can take long excursions in order to cover short distances on
the lattice. Indeed, the geometric constraints imposed by the lattice
are removed by the disorder, and the geodesics are never constrained
to follow the axis (such as in the Gaussian regime for $k>k^{\star}$ and
$x\ll d_c$) and explore the space freely in any direction.

Moreover, Fig. \ref{fig:passage_time_deviation_axis_Weibull} shows
that the KPZ regime, $\sigma_T\sim x^{1/3}$, is always recovered for
distances above a new crossover length, $x\gg \ell_o$, with $\ell_o$
increasing as $k$ decreases. In the figure we have indicated with
vertical segments and shadowed intervals theoretical estimates for the
location of this crossover length according to the calculations that
we present later. See Sec. \ref{sec:crossover} for an explanation and
discussion of these estimates.

A thorough explanation of the results presented in this section
will be given in the following sections.

\section{Mapping of FPP under extreme disorder into bond percolation}
\label{sec:percolation}

Strongly disordered link-time distributions lead to an interesting
phenomenon: the arrival time, which is by definition the sum of all
link-times along the geodesic path, is {\em dominated by its largest
  term}. Then, we can assert that
\beq
T(\bx)=\min_\Gamma \sum_{i=1}^m t(\bx_{i-1},\bx_{i})
\approx \min_\Gamma \left\{\max_i \{ t(\bx_{i-1},\bx_{i})\}\right\},
\label{eq:minmax}
\eeq
i.e. arrival times follow a min-max principle. It is not hard to see
that, if the max principle holds exactly, the set of points which can
be reached in time $T$ corresponds to the set of points for which
there is a path which never crosses a link with crossing-time larger
(or equal) than $T$. Moreover, the perimeter of the ball will be
surrounded by links with crossing times $t\geq T$. Thus, a clear
connection with percolation can be established: the FPP ball obtained
at time $T$, $B(T)$, corresponds to the cluster obtained in an
equivalent {\em bond percolation} problem in which open bonds
correspond to those FPP links with crossing-times lower than $T$,
whereas closed bonds are given by the FPP links with crossing-time
above $T$. To establish this equivalence we need to define the
probability $p$ of a bond being open ($1-p$ of being closed). Under
the assumption made in Eq. \eqref{eq:minmax} this probability is given
by the cumulative distribution function of the link-times evaluated at time
$T$:
\beq
p=F(T)=\int_0^T f(t)\,dt.
\label{eq:pFT}
\eeq
This section is devoted to providing strong evidences that under extreme
disorder conditions the FFP problem can be mapped into a
bond-percolation problem through relation \eqref{eq:pFT}, which is the
cornerstone of this work. Clearly the inverse transformation is given
by $T=F^{-1}(p)$. For example, for the Weibull distribution we obtain: \beq
\begin{aligned}
p(T)&=1- \exp\(-(T/\lambda)^k\),\\
T(p)&=\lambda \(-\ln \left[1-p\right]\)^{1/k}.
\end{aligned}
\label{eq:wei_p(T)}
\eeq
Notice that $p(T)$ increases continuously with $T$ from $p(0)=0$ up to
$p\rightarrow 1$ when $T\rightarrow \infty$. Also, the passage time
corresponding to a given probability decreases monotonically as $k$
decreases, approaching $0$ as $k\rightarrow 0$.

We will show that FPP balls with arrival time under $T$ are equivalent
to bond-percolation clusters obtained at bond probability
$p=F(T)$. The mapping given in Eq. \eqref{eq:pFT} will allow us to
obtain some of the most relevant critical exponents of the associated
percolation problem from the scaling analysis of the FPP geodesic
balls. In the analysis we will use the standard notation of
percolation theory \cite{Stauffer_03}. Since we are interested in the
strong disorder regime we have to consider very low values of the
order parameter $o\ll o^{\star}$. In the case of the Weibull
distribution we shall focus on three values: $k=0.03$, $0.01$ and
$0.005$, because they are sufficiently low to display the transition
to the percolation phase at a reasonable computational cost (we recall
that $k^{\star}\approx 1.79$).

\subsection{Ball size distribution}

Percolation theory \cite{Stauffer_03} deals primarily with the
statistical properties of the clusters of neighbouring sites which are
occupied (site-percolation) or connected by open bonds
(bond-percolation), when each site (or bond) is occupied (open) with a
probability $p$. The {\em critical point} or {\em percolation
  threshold} $p_c$ is defined as the minimal probability $p$ for which
an infinite percolation cluster is formed in an infinite lattice. Near
the critical point the system is characterized by a set of critical
exponents that are independent of the type of percolation or the
lattice geometry, depending only on the dimension $D$ of the
lattice. On the other hand, the percolation threshold varies with all
these factors. For bond-percolation in a $D=2$ square lattice we have
$p_c=1/2$.

Let $n_s(p)$ be the number of clusters of size $s$ (i.e. with $s$
sites), divided by the total number of lattice sites. This observable
is known to present the following scaling relation near criticality
($p\rightarrow p_c$) and for large clusters ($s\rightarrow \infty$):
\beq
n_s(p)=s^{-\tau} g\((p-p_c)s^\sigma\),
\label{eq:nsp}
\eeq
where $g(z)$ is a scaling function that approaches a constant value
for $|z|\ll 1$ and decays exponentially for $|z|\gg 1$. This scaling
relation introduces a crossover size $s_\xi$, which scales as
$s_\xi\sim|p-p_c|^{-1/\sigma}$, such that:
\beq
n_s(p) \sim s^{-\tau} \quad {\rm for} \;\; s\ll s_\xi.
\label{eq:nsp2}
\eeq
Beyond $s_\xi$, the power-law behavior fails and the number of
clusters decays fast. In $2D$ we have $\tau=187/91$ and
$\sigma=36/91$.

\begin{figure}
  \includegraphics[width=\columnwidth]{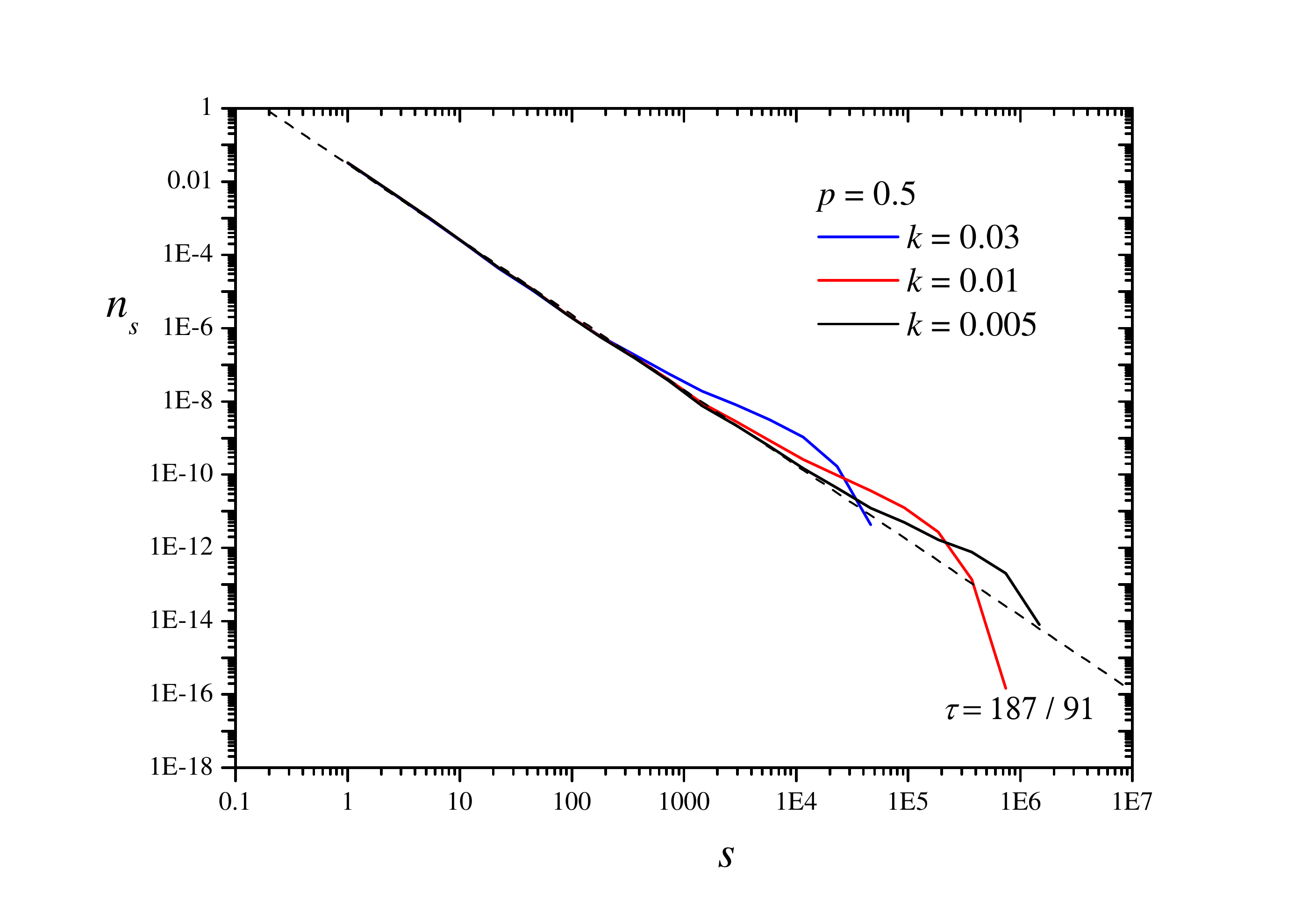}
  \caption{Ball size distribution, derived from the probability $W_s$
    of obtaining a FPP ball with size $s$, using Eq. \eqref{eq:wsns},
    for crossing-times following a distribution Wei(1,$k$) with
    $k=0.03$, $0.01$ and $0.05$, and for the arrival time $T$ which
    corresponds to $p=p_c=1/2$ according to
    Eq. \eqref{eq:wei_p(T)}. These arrival times are, respectively:
    $T=4.95\cdot 10^{-6}$, $T=1.21\cdot 10^{-16}$ and $T=1.46\cdot
    10^{-32}$. The broken line indicates the scaling given in
    Eq. \eqref{eq:nsp2}.}
  \label{fig:tau}
\end{figure}

In a FPP framework we do not have access to this number of
clusters. Yet, given a fixed arrival time $T$ we can estimate the
probability that the FPP ball contains exactly $s$ sites, denoted here
by $W_s(T)$. Since the center node always belongs to the FPP ball,
this value can be mapped, in the percolation language, to the
probability that a randomly chosen \emph{cluster} site will be part of
a cluster of size $s$. Let $w_s(p)$ be the probability (in the
percolation problem) that the cluster to which a randomly chosen
occupied site belongs has size $s$. It is easy to show that
$n_s(p)=p\, w_s(p)/s$. We can now relate $W_s(T)$ and $w_s(p)$ through
the mapping between arrival times and occupation probabilities given
in Eq. \eqref{eq:pFT}, so we have $w_s(p(T))=W_s (T)$. Thus, for a
given arrival time $T$, we propose that our FPP ball distribution is
given by
\beq
n_s\(p(T)\)={p(T)\, W_s(T)\over s}.
\label{eq:wsns}
\eeq

Our analysis of the statistics of the FPP balls begins in
Fig. \ref{fig:tau}, in which we show the ball size distribution
calculated from Eq. \eqref{eq:wsns} for the three distributions of
reference and for arrival times corresponding to the critical
probability $p_c=1/2$. We first observe a perfect collapse of the
three curves into the scaling law given in Eq. \eqref{eq:nsp2} over
several orders of magnitude. In bond percolation, the cutoff size
$s_\xi$ diverges at $p_c$, so we should expect a continuous power-law
decay. However, the plot shows that distributions deviate from the
naively predicted scaling after a given point, showing first an
increase with respect to the expected values leading to a final fast
decay. Let us denote this cutoff size by $s_o(p)$, where we have
assumed that, beyond its dependence on the order parameter $o$, it also depends on the probability $p$. It it clear that this crossover size increases as the disorder becomes stronger, which leads us to conjecture that it diverges as
$o\rightarrow 0$ for any $p$.

The crossover size $s_o$ appears as a consequence of the breakdown of
our main assumption: the max principle given in
Eq. \eqref{eq:minmax}. Let us associate a length scale to $s_o$, given
for example by the average radius of the balls of size $s_o$, and
denote it by $\xi_o$. For length scales of the order of $\xi_o$ the
geodesic paths are long enough to include links with crossing-times
that, while smaller than the largest term, do significantly contribute
to the arrival time. This means that, for a given arrival time $T$,
not all the links with a crossing-time lower than $T$ are allowed,
which leads to the breakdown of the max principle. As a result, the
clusters with sizes larger than $s_o$ predicted by percolation theory
are trimmed into FPP balls with smaller sizes, a fact that leads to
the increase of $n_s$ for $s>s_o$ displayed in the figure. We will
refer to this process as \emph{rounding-off}. For larger length
scales and thus longer geodesics ($s\gg s_o$), the round-off effect
limits the size of the distributed balls at a given arrival time $T$
and the distribution drops steeply.

To see clearly this effect we can consider the critical point in bond
percolation, where an infinite percolation cluster appears. Let us
define the arrival time corresponding to the percolation threshold as
the {\em critical arrival time} $T_c$: $T_c\equiv T(p_c)$. For the
Weibull distribution we obtain $T_c=\lambda (\ln2)^{1/k}$. An infinite
ball $B(T_c)$ would necessarily imply infinitely long geodesics with
an infinite number of links. Since link-times are positive we cannot observe an infinite ball for a finite arrival
time. Only for distributions with $F(0)=p_c$ (critical FPP) or
$F(0)>p_c$ (supercritical FPP) we observe infinite balls.

We can thus identify $\xi_o$ as the length scale characterizing the
crossover observed in the model: for length scales much smaller than
$\xi_o$ the behavior is the same as in bond percolation with
$p=F(T)$.

At length scales of the order of $\xi_o$, we may postulate the
existence of an equivalent bond-percolation problem characterized by
an effective probability smaller than $p=F(T)$, because not all links
with times below $T$ are allowed. The upper limit of the integral
given in Eq. \eqref{eq:pFT} can no longer be given by $T$, but by
$T-\epsilon_T$ with some $\epsilon_T>0$. We conjecture that this
results in an \emph{effective} critical probability $p_{c,\eff}$ which
is slightly larger than the theoretical percolation threshold
$p_c=1/2$. A rough argument for this is as follows. Let us consider
the inverse problem and fix the occupation probability $p$ in the
bond-percolation problem. Following the above reasoning we should
expect that the effective arrival time $T_{\eff}$ of the equivalent
FPP problem should be of the form $T_\eff(p)=T(p)+\epsilon_T(p)$, with
$T(p)=F^{-1}(p)$ and some $\epsilon_T(p)>0$. At the critical point
$p_c$ we can then write
$T_{c,\eff}(p_c)=T(p_c)+\epsilon_T(p_c)=T_c+\epsilon_T(p_c)$. We can
now define the \emph{effective} critical probability $p_{c,\eff}$ as
the probability that satisfies $T(p_{c,\eff})=T_{c,\eff}(p_c)$. We
then have $p_{c,\eff}=F(T_{c,\eff}(p_c))=
\int_{0}^{T_{c,\eff}(p_c)}f(t)dt=
\int_{0}^{T_c+\epsilon_T(p_c)}f(t)dt=p_c+\epsilon_p$, where
$\epsilon_p=\int_{T_c}^{T_c+\epsilon_T(p_c)}f(t)dt>0$.

It is expected that $p_{c,\eff}$ will depend on the order
parameter. To obtain the effective critical probability for each $k$
we have represented the ball size distribution $n_s(p)$ obtained at
different values of $p(T)$ and we have identified the value of $p$ at
which the cutoff value $s_o(p)$ where the ball distribution departs
from the theoretical scaling law takes its maximum value
(i.e. $p_{c,\eff}$ yields the largest value of $s$ for which $n_s$
still lies on the straight line). For case $k=0.03$ we have observed
that the maximum cutoff value is obtained at $p\approx 0.505$ so we
have $p_{c,\eff}(0.03)\approx0.505$. Repeating the procedure for cases
$k=0.01$ and $k=0.005$ we obtain $p_{c,\eff}(0.01)\approx 0.502$ and
$p_{c,\eff}(0.005)\approx 0.501$.

Although these values are far from being accurate, the collapses
displayed in the next subsection support our claims. It is very common
in percolation studies to consider an effective critical probability
different from the theoretical value in order to take into account
finite-size effects. In our model these finite size effects appear as
a consequence of the fact that the max principle given in
Eq. \eqref{eq:minmax} does not hold strictly. We expect that
$p_{c,\eff}(k)\rightarrow p_c$ as $o\rightarrow 0$.

\subsection{Average cluster size}

In percolation theory the average cluster size $\<s(p)\>$ is usually
defined \cite{Stauffer_03} as the first moment of the size
distribution obtained from the random selection
of some cluster site (defined above as $w_s(p)$). We then have
$\<s(p)\>=\sum_s s\,w_s(p)$. As we approach criticality from below
($p\rightarrow p_c^-$) the average cluster size diverges according to
the following scaling law:
\beq
\<s(p)\> \sim |p_c-p|^{-\gamma},
\label{eq:gamma}
\eeq
with the critical exponent $\gamma=43/18$ for $D=2$. This relation
also holds when we approach the critical point from above
($p\rightarrow p_c^+$) provided that we exclude the single infinite cluster
in the sum over all cluster sizes.

\begin{figure}
  \includegraphics[width=8cm]{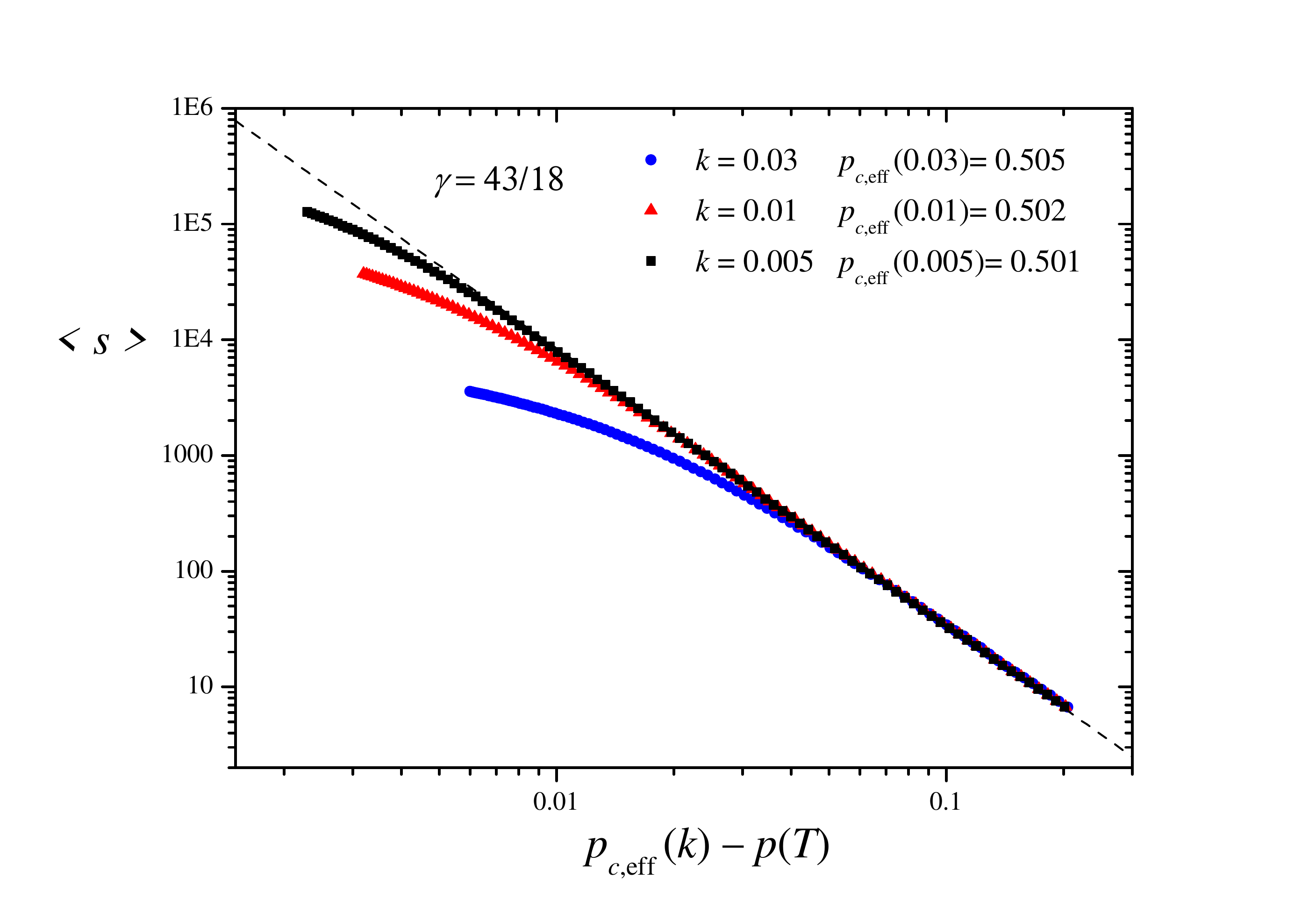}
  \caption{Average cluster size of the FPP balls calculated from
    Eq. \eqref{eq:averages} as a function of $p_{c,\eff}(k) - p(T)$
    for the three reference distributions. Corresponding
    $p_{c,\eff}(k)$ are shown in the legend. Straight broken line
    corresponds to the theoretical scaling given in
    Eq. \eqref{eq:gamma}.}
  \label{fig:gamma}
\end{figure}

We can then calculate the average FPP-ball size in a similar way:
\beq
\<s(p(T))\>=\sum_s s\,W_s(T).
\label{eq:averages}
\eeq
Numerical evidence that the balls of the FPP model also obey the above
scaling is shown in Fig. \ref{fig:gamma}, where we have used the
effective critical probabilities discussed above. For values of $p$
much smaller than the critical value we observe an excellent collapse
of the three curves into the expected power-law. When $p$ is close to
the critical point the main contribution to the sum comes from large
values of $s$. In percolation theory the size of these clusters is
$s_{\xi}$, which diverges at the critical point. In our model we have
found another crossover size, $s_o$, above which the round-off
prevents balls with larger sizes. The interplay between both
characteristic sizes can explain the behavior obtained in our
simulations. For probabilities such that $s_{\xi}(p) \ll s_o(p)$ the
round-off does not affect the dominating size $s_{\xi}(p)$ and
FPP-balls scale as percolation clusters. As $p$ approaches the
critical point, $s_{\xi}$ diverges becoming larger than $s_o$, which
now turns into the size that dominates the moments of the mass
distribution. This yields smaller average sizes and thus a deviation
with respect to the theoretical scaling. The crossover probability
$p_o^{\star}$ (crossover arrival time $T_o^{\star}$) at which it
occurs is thus obtained when $s_{\xi}(p_o^{\star})$ is of the same
order as $s_o(p_o^{\star})$. In the previous section we obtained that
$s_o$ increases as the disorder becomes stronger, so $p_o^{\star}$ should also increase as $o$ decreases, in agreement with the behavior displayed in the figure. We can thus expect that $p_o^{\star} \rightarrow p_c$ as
$o\rightarrow0$.

\subsection{Correlation length}

Critical behavior in percolation theory is completely dominated by a
single characteristic length, the \emph{correlation length} $\xi$
\cite{Stauffer_03}, which stands for the crossover length scale below
which the behavior is indistinguishable from that at $p_c$.

Given a single percolation cluster, its \emph{radius of gyration} is
defined by the average squared distance between two cluster sites:
\beq
R_s^2\equiv {1\over 2} \sum_{i,j} {|\textbf{r}_i-\textbf{r}_j|^2\over s^2},
\label{eq:radius}
\eeq
where $\textbf{r}_i$ stands for the position vector of the $i$-th site
of the cluster and the subscript $s$ stands for the cluster size. We
can average this radius over all clusters of size $s$ to obtain
$\overline{R_s^2}$. Finally, the average of $\overline{R_s^2}$ over
all cluster sizes $s$ in the following way provides a standard
definition of the squared correlation length \cite{Stauffer_03}:
\beq
\xi^2(p) \equiv 2 {\sum_s \overline{R_s^2} s^2 n_s(p)\over \sum_s s^2 n_s(p)},
\label{eq:xi_percolation}
\eeq
which is known to diverge as we approach the critical point
($p\rightarrow p_c$) as
\beq
\xi(p)\sim |p_c-p|^{-\nu},
\label{eq:nu}
\eeq
with $\nu=4/3$ for $D=2$. Again, if we approach from above we have to
exclude in the sum the contribution of the infinite cluster.

The correlation length is the radius of those clusters which mainly
contribute to the second moment of the cluster size distribution. Near
the percolation threshold this contribution comes from the clusters of
size of the order of $s_{\xi}$ so we find
\beq
s_{\xi}\sim \xi^{D_f},
\label{eq:s_xi}
\eeq
where $D_f$ is the fractal dimension of the infinite percolation
cluster, $D_f=91/48$ for $D=2$.

\begin{figure}
  \includegraphics[width=8cm]{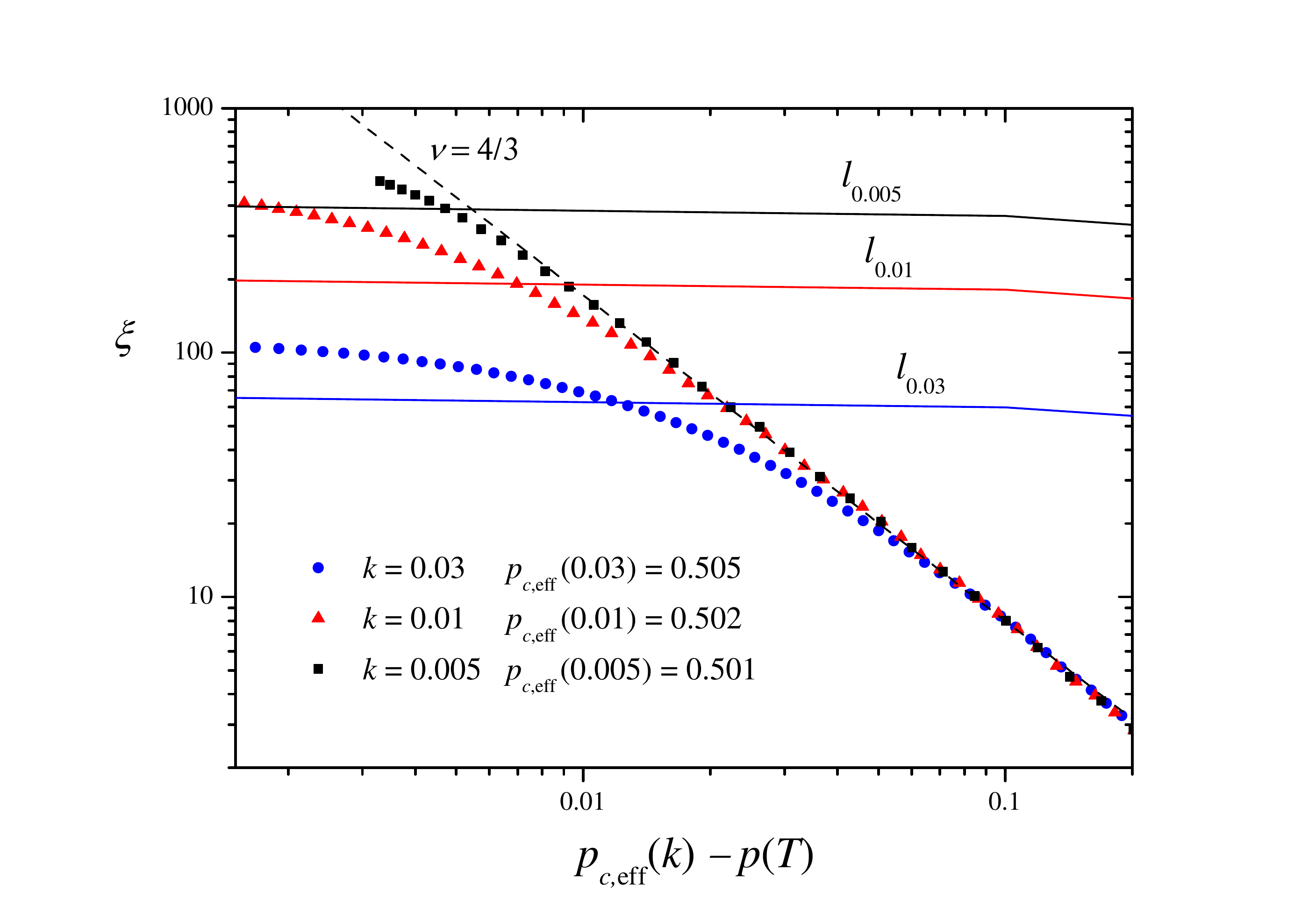}
  \caption{Correlation length of the FPP balls calculated from
    Eq. \eqref{eq:xi_FPP} as a function of $p_{c,\eff}(k) - p(T)$ for
    the three reference distributions. Corresponding $p_{c,\eff}(k)$
    values are shown in the key. Straight broken line corresponds to
    the theoretical scaling given in Eq. \eqref{eq:nu}. Continuous
    curves correspond to the length $\ell_o(T(p))$ obtained for each
    $k$ from the chain model presented in Sec. \ref{sec:chain_model}.}
  \label{fig:nu}
\end{figure}

If we turn now to our FPP problem and apply the equivalence discussed
so far to Eq. \eqref{eq:xi_percolation}, we deduce the following
expression for the correlation length of the FPP balls:
\beq
\xi^2(p(T))=2{\sum_s \overline{R_s^2}\, s W_s(T)\over \sum_s s W_s(T)}.
\label{eq:xi_FPP}
\eeq
Results regarding the correlation length have been displayed in
Fig. \ref{fig:nu} following the same scheme as in Fig. \ref{fig:gamma}
for the average size. The similarity between both plots is not
surprising since they correspond to moments of the same cluster size
distribution. According to percolation theory, $s_{\xi}$ is exactly
the cluster size that dominates the moments of the mass distribution,
including the average cluster size $\<s\>$ and the correlation length
$\xi$. As a consequence, $\xi$ represents the radius of the clusters
of size $s_{\xi}$. On the other hand, the radius $\xi_o$ of balls of
size $s_o$ gives the crossover length scale below which the balls
behave as percolation clusters. We can thus use here the explanation
given for the average cluster size by considering the correlation
length $\xi$ instead of $s_{\xi}$, and the cutoff radius $\xi_o$
instead of $s_o$. For probabilities such that $\xi(p) \ll \xi_o(p)$,
the round-off effect is negligible since it only affects sizes much
larger than the dominant size $s_{\xi}$. Thus, the correlation length
of the FPP balls behaves as in the percolation problem. As we approach
the critical point, $\xi$ diverges, becoming much larger than
$\xi_o$. Due to the round-off effect size $s_o$ becomes the dominant
term, yielding a deviation with respect to the scaling law. The
crossover probability $p_o^{\star}$ is the same as for the average
cluster size and is obtained when $\xi(p_o^{\star})$ is of the same
order as $\xi_o(p_o^{\star})$.

\subsection{Percolation probability}

The order parameter of the second-order phase transition observed in percolation is given by the \emph{percolation probability}
$\c{P}$, or \emph{strength of the infinite cluster}, defined as the
probability that a randomly chosen site belongs to the infinite
cluster. For $p\leq p_c$ we have $\c{P}=0$, and it goes to zero as a
power law as we approach the critical point from above ($p\rightarrow
p_c^+$):
\beq \c{P}(p)\sim (p-p_c)^\beta,
\label{eq:beta}
\eeq
where $\beta=5/36$ for $D=2$ is yet another critical exponent
\cite{Stauffer_03}. Our purpose is to recover this exponent from the
analysis of the FPP balls.

\begin{figure}
  \includegraphics[width=9.5cm]{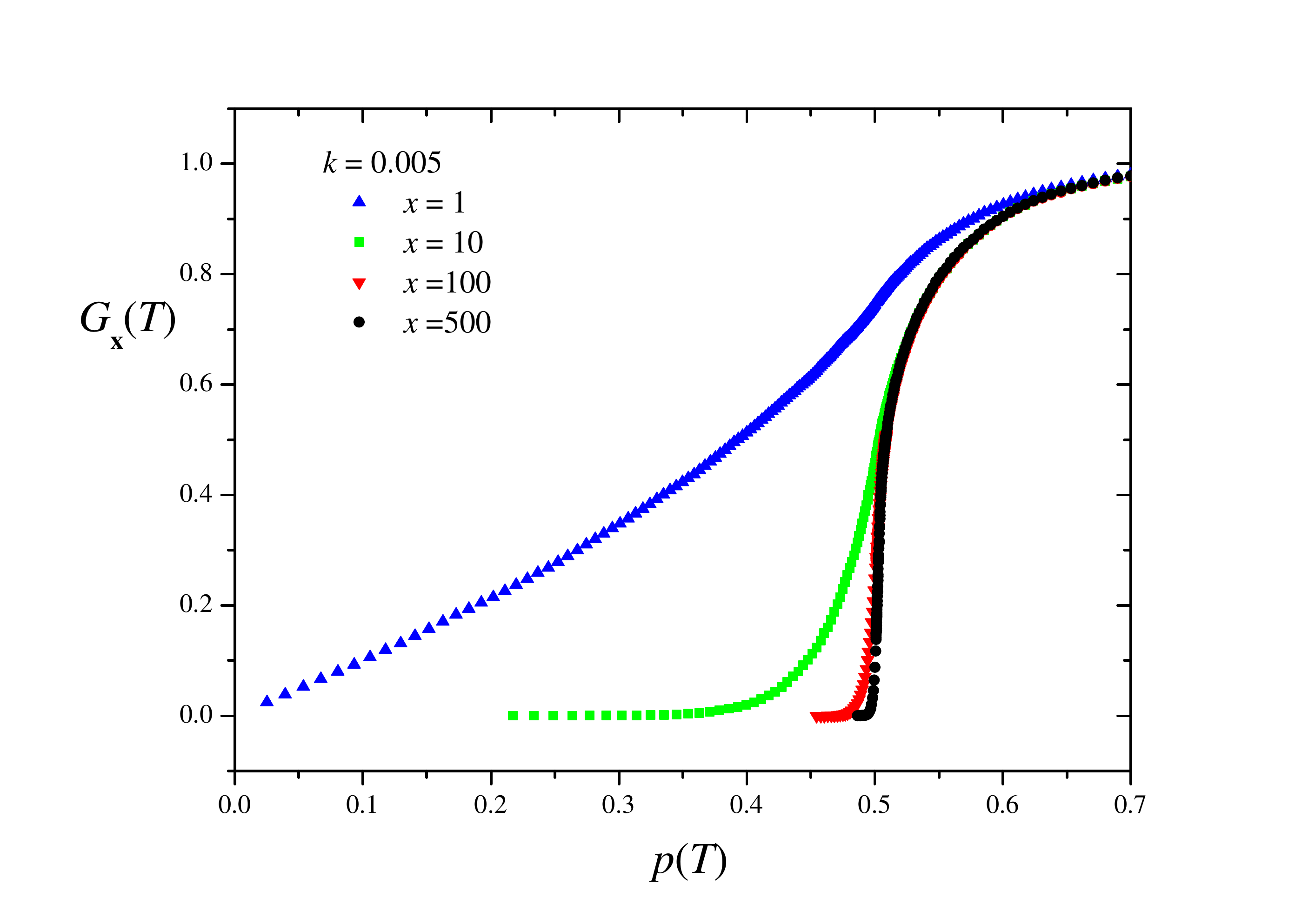}
  \includegraphics[width=9.5cm]{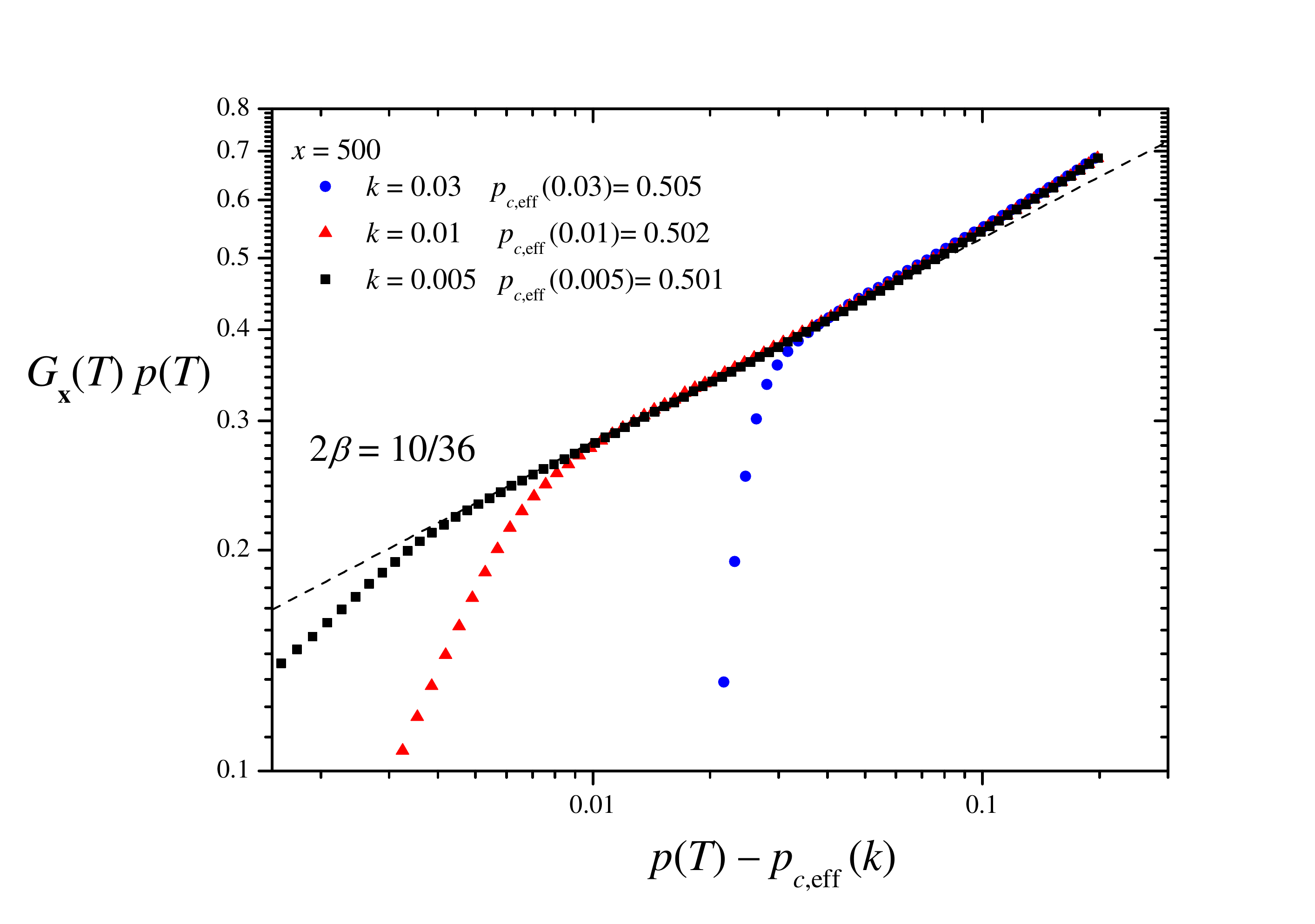}
  \caption{Strength of the infinite cluster. (Top) Probability that a
    node at position $x$ in the axis will belong to the ball with
    passage time $T$, as a function of $p=F(T)$ for $k=0.005$ and
    different distances: $x=1$, $10$, $100$ and $500$. (Bottom)
    $G_\bx(T)\,p(T)$ product as a function of $p(T)-p_{c,\eff}(k)$ for a
    node at $x=500$ and the three reference distributions. The
    straight broken line represents the scaling behavior deduced in
    Eq. \eqref{eq:FP}.}
  \label{fig:beta}
\end{figure}

Let us define $G_\bx(T)$ as the probability that site
$\mathbf{x}$ belongs to the FPP ball at passage time $T$,
\beq
G_\bx(T) = \text{Prob}\{\bx\in B(T)\}.
\label{eq:Fxt}
\eeq
$G_\bx(T)$ is also the probability that the arrival time needed to
reach site $\mathbf{x}$ is less than $\emph{T}$, i.e. the cumulative
distribution of the arrival time to node $\mathbf{x}$ at time $T$. Note that $G_\bx(T)$ is related to the probability density function
$g_\bx(T)$ addressed in previous section and displayed
in Fig. \ref{fig:histograma_T_Weibull}, through
$g_\bx(T)=dG_\bx(T)/dT$. The behavior of $G_\bx(T)$ as a function of
$p=F(T)$ for different sites along the axis for $k=0.005$ is
shown in the top panel of Fig. \ref{fig:beta}.

As we move away from the center node the curves become sharper around
$p=1/2$ and, for large $x$, $G_\bx(T)$ tends to a limit function. That
limit function is qualitatively very similar to the plot of the order
parameter $\c{P}$ as a function of $p$. For distant positions there is
a minimum arrival time, which approaches very fast the critical
passage time $T_c$ as $x \rightarrow \infty$. This result is
consistent with the criticality of the percolation problem: at $p=p_c$
an infinite cluster that percolates through the lattice appears for
the first time.

From these results and Eq. \eqref{eq:Fxt} we deduce that for
$x\gg 1$, $G_\bx(T)$ is the probability that both the initial
and the final points belong to the infinite cluster. In percolation
theory this probability is given, for very long distances (and thus
uncorrelated sites) by $\c{P}^2$. Besides, this probability is
conditioned on the known fact that the initial point (the center node) belongs to the cluster, so we have to divide the above expression by $p$. Our
reasoning allow us to conjecture that for $x \gg 1$:
\beq
G_{\bx}(T) \approx {\c{P}^2(p) \over p}\rightarrow
{(p-p_c)^{2\beta}\over p} \quad \mbox{as} \quad p\rightarrow p_c,
\label{eq:FP}
\eeq
using always $p=F(T)$. Numerical evidence for Eq. \eqref{eq:FP} is
shown in the bottom panel of Fig. \ref{fig:beta}. The plot displays
the values of the product $p\, G_{\bx}(T)$, obtained from the top
panel of the figure, as a function of $p-p_{c,\eff}(k)$, for the three
levels of disorder. As in previous figures we obtain an excellent
collapse of the curves into the expected power-law as $p\rightarrow
p_{c,\eff}(k)$, while deviations take place at the same crossover
probabilities $p_0^{\star}$ obtained before for the mean cluster size
and the correlation length.

\subsection{Fractal dimension of FPP balls}

The infinite cluster at the critical point is a fractal, so its mass
$M$ scales with the linear size $\ell$ as $M\sim \ell^{D_f}$, where
$D_f$ is the \emph{fractal dimension} ($D_f=91/48$ for $2D$). However,
fractal behavior is also observed away from the critical point at
length scales much smaller than the correlation length, so we have
$M\sim \ell^{D_f}$ for $\ell \ll \xi$, and $M\sim \ell^{D'}$ for $\ell
\gg \xi$, with dimension $D'$ depending on whether we are above or
below the critical point (for $p>p_c$ we have $D'=D$, the Euclidean
dimension). Clusters look fractal on scales smaller than $\xi$ and
this also applies to the relation between their mass $s$ and their
linear size $R_s$: for $R_s \ll \xi$ or equivalently $s \ll s_{\xi}$
we find $s\sim R_s^{D_f}$, from which we can deduce
Eq. \eqref{eq:s_xi}.

\begin{figure}
  \includegraphics[width=9.5cm]{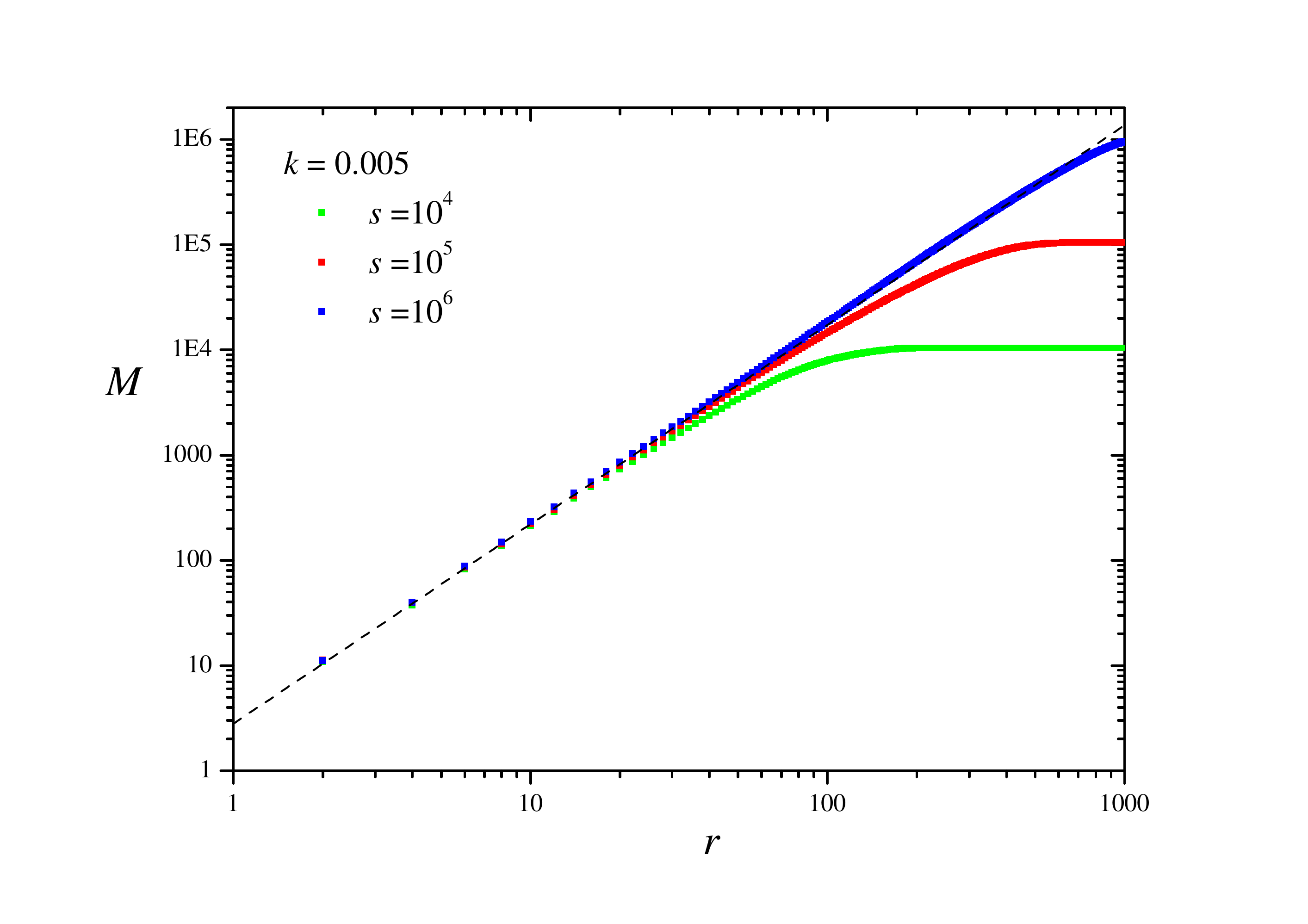}
  \caption{Mass of the FPP ball $B(T_c)$ within a circle of radius $r$
    centered at the origin for case $k=0.005$. Results correspond to
    the average over balls with sizes within the following intervals:
    (green) $s\in(1,1.1)\cdot 10^4$, (red) $s\in(1,1.1)\cdot 10^5$ and
    (blue) $s\in(1,1.1)\cdot 10^6$. The straight line corresponds with
    a power-law, $M\sim r^{91/48}$.}
  \label{fig:massdim}
\end{figure}

In Fig. \ref{fig:massdim} we have shown the mass of FPP balls within
circles of increasing radius $r$ centered at the origin, for extreme
disorder $k=0.005$. To perform the analysis we have selected among all
the FPP balls grown at critical passage time $T_c$ those with sizes
contained in small intervals around three different sizes $s$. Results
were averaged over all the balls within the same interval.

The plots show a constant slope which coincides with the fractal
dimension for critical percolation (broken line). The final plateaux
are not related to a crossover. They are due to the fact that clusters
become finite with linear size $R_s$, so after $r\approx R_s$ the mass
becomes constant. It is interesting to note that sizes $s=10^5$ and
$s=10^6$ are larger than the corresponding cutoff size $s_o$ (above
$10^4$, see Fig. \ref{fig:tau}), i.e. these clusters are influenced by
the round-off. However, this effect does not seem to significantly
change the internal structure of the balls, which is still fractal, so
the main consequence of the round-off is the trimming of the branched
edges.

\section{Crossover length for percolation in the FPP model}
\label{sec:chain_model}

Let us elaborate on the validity of the max principle that allows the
mapping of the FPP into the percolation problem. Our purpose is to
find an estimate for the cutoff values $\xi_o(p)$ and $s_o(p)$.

Another way to see the max principle is presented in the following
experiment, which we call the \emph{chain model}. Let us consider a
linear array of $\ell$ sites and let us choose a time $T$. Then we
fill the chain with $t$ values obtained by sampling the link-time
distribution, but only keeping values which are smaller than $T$,
i.e. if $t\geq T$ we disregard the value. These chains represent
idealized versions of the minimal paths since they cannot include
crossing-times larger than the passage time $T$. We define $\mathbf{P}$ as
the probability for the sum of the link-times to be smaller than $T$:
\beq
\mathbf{P}_o(\ell,T) =
\text{Prob}\left\{\sum_{i=1}^{\ell} t_i<T \; | \; t_i<T\right\}.
\label{eq:Pmodel}
\eeq
$\mathbf{P}_o(\ell,T)$ can be seen as the probability that the max
principle holds at a length scale $\ell$ and passage time $T$. Our
claim is the following: if there exists a cutoff length $\ell_o(T)$
such that
\beq
\mathbf{P}_o(\ell,T) \approx \begin{cases} 1 & \textrm{for } \ell\ll \ell_o(T), \\
 \textrm{fast decay} & \textrm{for }\ell\gg \ell_o(T), \end{cases}
\label{eq:ell}
\eeq
then $\xi_o(p)$ should behave as $\ell_o(T)$, always with
$p=F(T)$. The numerical evidences that follow will support our claim.

\begin{figure}
  \includegraphics[width=8cm]{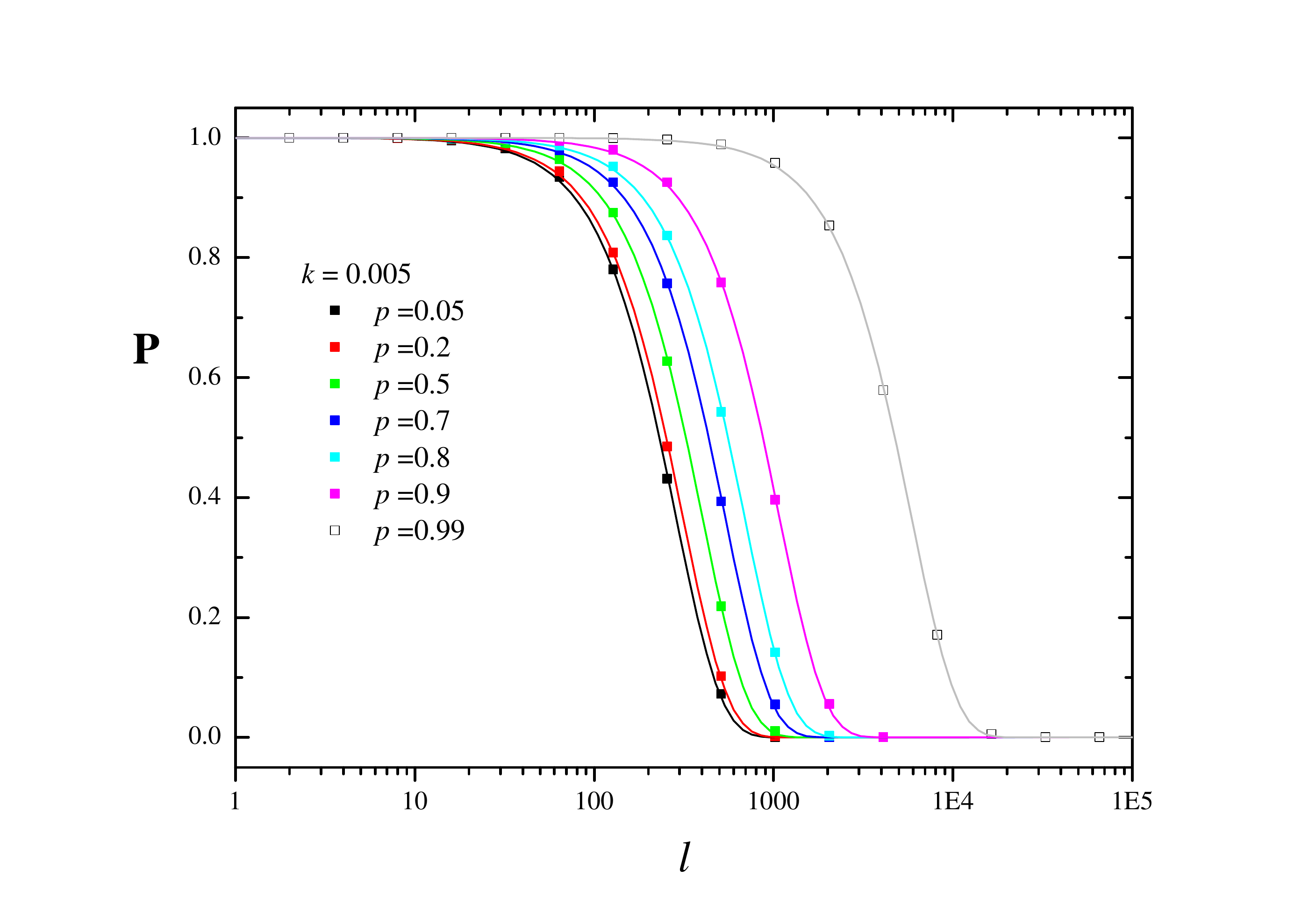}
  \includegraphics[width=8cm]{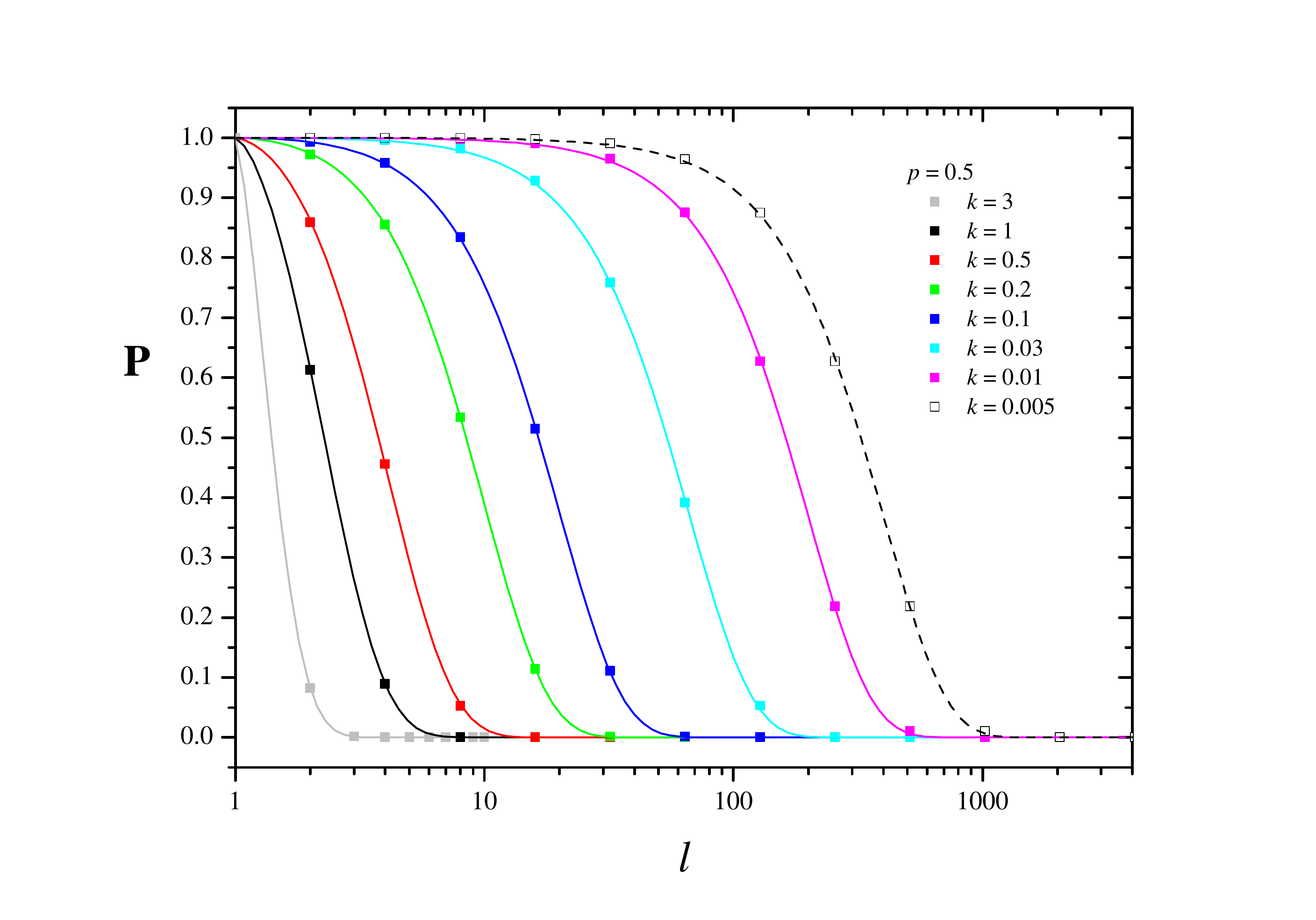}
  \caption{Probability $\mathbf{P}$ calculated from
    Eq. \eqref{eq:Pmodel} over an ensemble of $10^6$ chains as a
    function of the chain length $\ell$ for: (top) $k=0.005$ and
    different passage times represented by $p(T)$; (bottom) $p=0.5$
    and different $k$. Points correspond to results from simulations
    while continuous lines represent the corresponding fits to
    function in Eq. \eqref{eq:CEF}.}
  \label{fig:P}
\end{figure}

Figure \ref{fig:P} shows two sets of results for
$\mathbf{P}_o(\ell,T)$. In the top panel we display the dependence of
$\mathbf{P}$ on $\ell$ for different arrival times $T$ (expressed as
$p(T)$) and for fixed $k=0.005$. In the bottom panel the arrival time
has been fixed to $T_c$ and the curves correspond to different
strengths of the disorder (different values of $k$).

The first remarkable
result is that $\mathbf{P}_o(\ell,T)$ shows an excellent agreement
with the \emph{stretched exponential function}:
\beq
\mathbf{P}_o(\ell,T) \approx
\exp\left[-\(\frac{\ell-1}{\ell_o(T)}\)^{\phi_o(T)}\right].
\label{eq:CEF}
\eeq
The corresponding fits have been shown with colored continuous
lines. As shown, the dependence of $\mathbf{P}$ on $\ell$ follows the
behavior conjectured in Eq. \eqref{eq:ell} with $\ell_o(T)$ thus
playing the role of a cutoff length. It is important to stress that we
have obtained the same behaviour for the other link-time distributions
investigated in this work (Log-Normal and Pareto).

\begin{figure}
  \includegraphics[width=8cm]{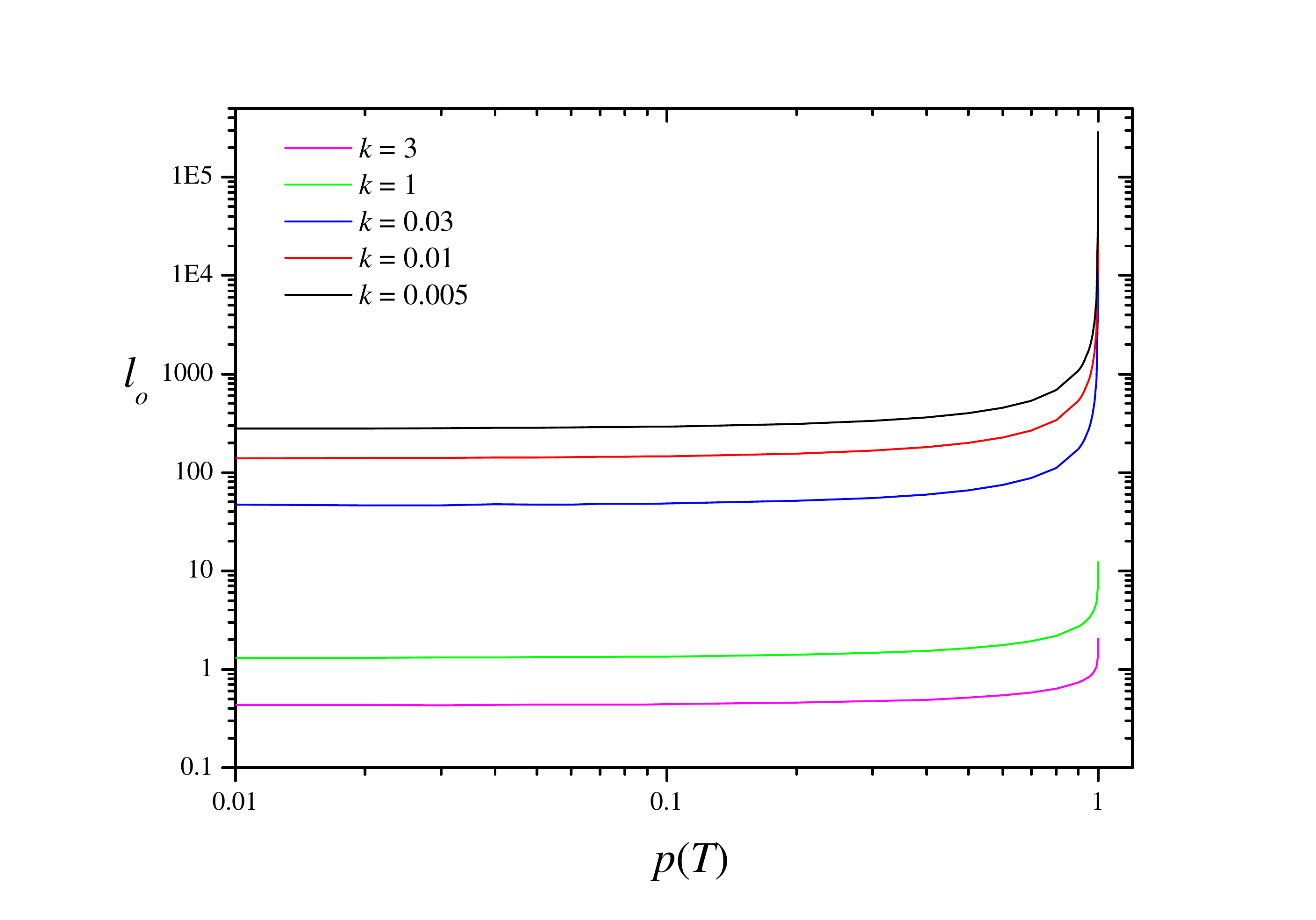}
  \includegraphics[width=8cm]{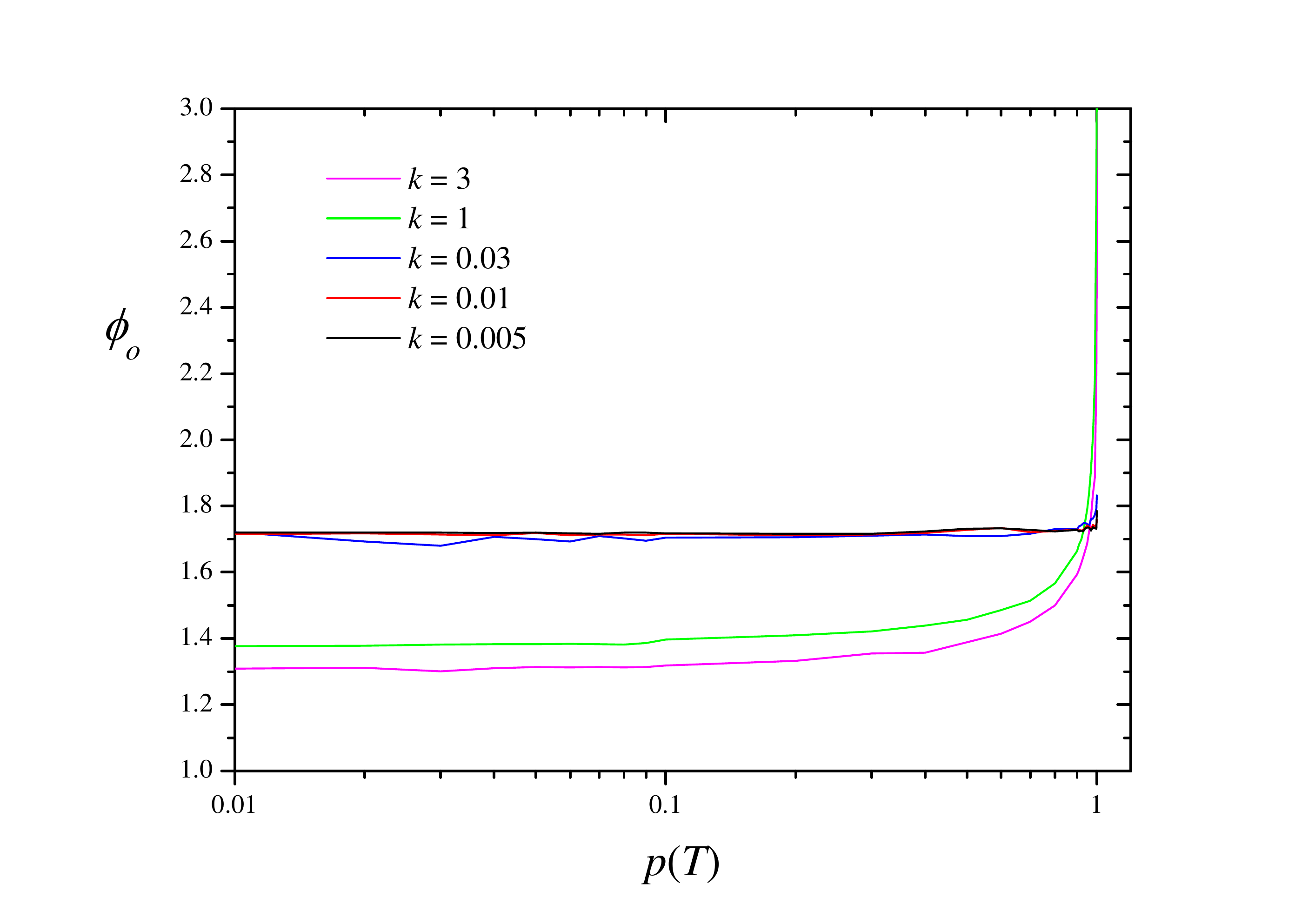}
  \caption{Fitted values of the cutoff length $\ell_o(T)$ (top) and
    the stretch exponent $\phi_o(T)$ (bottom) as a function of
    probability for different strengths of disorder, $p(T)$.}
  \label{fig:lk}
\end{figure}

We address now the behavior of the fitting parameters, the cutoff
length $\ell_o(T)$ and the stretch exponent $\phi_o(T)$, displayed in
the top and bottom panels of Fig. \ref{fig:lk} respectively, as a function
of the probability $p(T)$ for different values of $k$.

With regard to $\ell_o(T)$, our results support all the assumptions made
in the previous section about $\xi_o$. For a given arrival time it increases
monotonously with the disorder, and for a given disorder it is a
continuously increasing function of the arrival time. We expect
$\ell_o(T)$ to diverge for any arrival time as $o\rightarrow
0$. Interestingly, although $\ell_o(T)$ is not defined at $T=0$ it
shows a well-defined positive limit. The same was found for the
Log-Normal distribution while for Pareto the limit is $0$.

Another remarkable result is the change of behavior near the crossover
point between weak and strong disorder ($k^{\star}=1.79$). For
$k<k^{\star}$ we obtain $\ell_o(T)>1$ for all $T$, in consistency with
our assumption that under strong disorder conditions there is always a
length scale below which the FPP model behaves as a percolation
lattice. However, if $k>k^{\star}$ (see e.g. case $k=3$) the fitted
values of $\ell_o(T)$ are smaller than 1 for $p<1$. Indeed, as $k$
increases above $k^{\star}$ the decay of $\mathbf{P}_o(\ell,T)$ after
$\ell=1$ becomes sharper and the expression given in
Eq. \eqref{eq:ell} approaches the step function:
$\mathbf{P}_o(\ell,T)=1$ if $\ell=1$ and $\mathbf{P}_o(\ell,T)=0$ if
$\ell>1$. As a consequence, the fitted values of $\ell_o(T)$ approach
$0$ as $k$ increases for finite $T$ (or $p<1$). We can then expect
that the function $\ell_o(p)-1$ approaches the delta function
$\delta(1-p)$ as $k\rightarrow \infty$. We will return to this issue
when we address the homogeneous case. Nevertheless, the change of
behavior of $\ell_o(T)$ near the crossover point $o^{\star}$ points to
it as a reliable estimate for the crossover length
$\xi_o$.

With regard to the stretch exponent $\phi_o(T)$, displayed in the
bottom panel of Fig. \ref{fig:lk}, we observe that for a given
$p$ it increases with disorder and seems to approach a
constant value above $1.7$ under strong disorder conditions. We also
notice that, in this regime, it roughly remains constant with
$p$. Only for large $k$ we observe a smooth increase
with passage time that sharpens as we approach $p=1$.

In order to check the relation between the cutoff length $\ell_o$
obtained from our chain model and the crossover length $\xi_o$
postulated as the length scale under which the FPP model behaves as a
percolation lattice, we have represented the corresponding curves of
$\ell_o$ in the plot of the correlation length $\xi$ of the FPP balls,
Fig. \ref{fig:nu}. The comparison between both magnitudes provide
numerical evidence of our claims. For probabilities $p$ such as $\xi
\ll \ell_o$, the correlation length of the balls follows the scaling
behavior predicted from percolation theory. As $p$ increases and $\xi$
approaches $\ell_o$, the effect of the round-off makes $\xi$ depart
from the percolation prediction. As discussed, this deviation takes
place at $p_o^{\star}$, where both quantities become of the same
order. A crossover towards a new behavior that its controlled by
$\ell_o$ takes place, in which we expect that $\xi$ will grow as $\ell_o$ for $p\gg
p_o^{\star}$.

\begin{figure}
  \includegraphics[width=8cm]{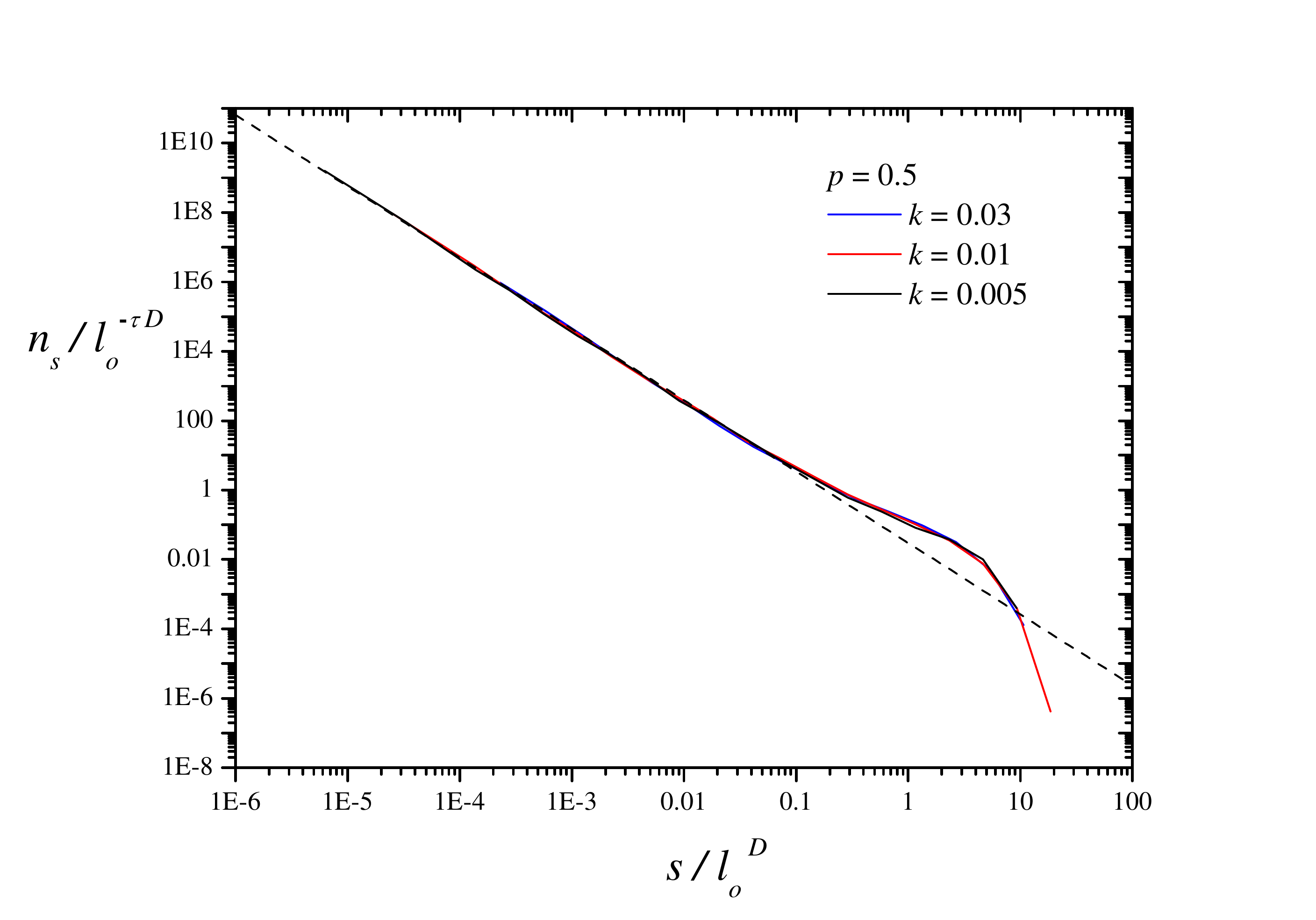}
  \caption{Rescaling of the size distributions displayed in
    Fig. \ref{fig:tau}. For each curve,
    ball size $s$ has been rescaled by $\ell_o^D$ ($D=2$) with values:
    $\ell_{0.03}(p=0.5)=65.83$, $\ell_{0.01}(p=0.5)=199.34$ and
    $\ell_{0.005}(p=0.5)=399.72$, obtained from the simulations of the
    chain model. $n_s$ has been rescaled by $\ell_o^{-\tau D}$
    according to Eq. \eqref{eq:nsk}.}
  \label{fig:collapse}
\end{figure}

We can provide further numerical evidence of the reliability of
$\ell_o$ as an estimator for the crossover length $\xi_o$. For
example, following our reasoning we can conjecture that $\ell_o$ is
the radius of the balls with cutoff-size $s_o$, so we can expect:
\beq
s_o\sim \ell_o^D.
\label{eq:skell}
\eeq
Then, according to Eq. \eqref{eq:nsp2} the number of clusters of size $s_o$, $n_{s_o}$, must behave as:
\beq
n_{s_o} \sim s_o^{-\tau} \sim \ell_o^{-\tau D}.
\label{eq:nsk}
\eeq
For probabilities close to the percolation threshold so that $s_o \ll
s_\xi$, the size distributions obtained from different $k$ should collapse
into a single universal curve if we rescale the size $s$ by $s_o$,
and $n_s$ by $n_{s_o}$. The rescaling of the distributions displayed in Fig. \ref{fig:tau} has been shown in Fig. \ref{fig:collapse}, and a quite satisfactory collapse of the three distributions is obtained.

\section{Discussion: crossover between percolation and KPZ scaling}
\label{sec:crossover}

\subsection{Strong disorder regime $o\ll o^{\star}$}
\label{subsec:strong}

The growth of FPP balls is controlled by two characteristic lengths:
the crossover length $\xi_o$ which determines the length scale below
which the FPP model is essentially a bond-percolation lattice with
$p=F(T)$, and the correlation length $\xi$ intrinsic to the
percolation problem. Whereas the behavior of $\xi$ is well-known from
percolation theory --at least close to the critical point--, we can
assume that the crossover length $\xi_o$ is related to the cutoff
length $\ell_o$ of the chain model. This is a strong assumption
because the link-times traversed by actual geodesics are correlated to
each other by means of the minimal time principle, in contrast to the
uncorrelated link-times assumed in our model.

For some distributions such as Weibull or Log-Normal, $\ell_o(T)$ has
a well-defined positive limit when $T\rightarrow 0$, so we can expect
the same for $\xi_o$. Since $\xi$ is initially $0$, we conclude that
FPP balls will display initial percolation-like growth provided this
limit value is larger than 1. This seems to be guaranteed in the
strong disorder regime $o<o^{\star}$. For the Pareto distribution we
find $\ell_o(T)\rightarrow 0$ as $T\rightarrow 0$. However, the
increase of $\ell_o(T)$ with passage time $T$ is faster than the
increase of $\xi(p)$ with probability $p(T)$, hence allowing an
initial percolation phase also in this case.

Percolation-like growth will take place whenever $\xi$ is much smaller
than $\xi_o$. In this regime, the growth of FPP balls as the arrival time
$T$ increases can be mapped into the growth of the percolation
clusters at increasing occupation probability $p=F(T)$.  That regime
continues until $\xi$ becomes of the same order as $\xi_o$, which
takes place at a certain arrival time $T_o^{\star}$. This
crossover time is upper bounded by the critical passage
time, i.e. $T_o^{\star}<T_c$, because $\xi$ diverges at the critical point. Under extreme disorder conditions, i.e. for
small but positive values of the order parameter $o$, $\xi_o$ is large, $T_o^{\star}$ is
close to $T_{c}$, and we can observe the fractal clusters obtained in
percolation. We expect that $T_o^{\star}\rightarrow T_c$ as
$o\rightarrow 0$, i.e. for infinite disorder.

Once the crossover arrival time $T_o^{\star}$ has been reached, we
enter into a transient regime that evolves towards KPZ scaling. Two
different types of growth at two different length scales take place
simultaneously. At scales below $\xi_o$, percolation-like growth still
continues with a percolation correlation length $\xi$ that rapidly
decreases with time above $T_c$ ($p> p_c$). The increase of $p$
fills the inner holes and cavities, and smoothes the irregularities of
the outer perimeter. Above $T_c$, the cutoff length $\xi_o$ prevents
the incipient infinite percolation cluster from spreading along the
lattice. At larger scales the growth of the ball is controlled by the
growth of $\xi_o$, which simply reflects the fact that minimal paths
are limited by their length. FPP balls becomes compact with a regular
shape, a process that finally leads to KPZ scaling.

This transient regime would last until the effects of percolation
vanish, i.e. when $p=1$, which corresponds to an infinite arrival
time. We can make, however, some speculations. For example, in
Fig. \ref{fig:passage_time_deviation_axis_Weibull} we showed the
fluctuations of the arrival time to points along the axis for
different levels of disorder. Now we know that $k=0.5$, $0.3$ or $0.2$
are not small enough to display the criticality of percolation, but
the disorder is in fact strong enough to reveal percolation effects in
the behavior of the arrival-time fluctuations. In each curve of the figure with
$k<k^{\star}$ we have marked three representative lengths obtained from the
chain model. As an approximation of $\xi_o$ we have considered
$\ell_o(T_{c})$, indicated with the vertical segments. In order to
estimate the range of the percolation effects and, thus, the crossover
towards KPZ scaling, we have represented in each curve with a grey
rectangle the interval $(\ell_o(T(p=0.99)),
\ell_o(T(p=0.999))$. Despite the approximations and the simplicity of
our model, the crossover points are in a reasonably good agreement
with the behavior displayed by the fluctuations.

Both the crossover length $\xi_o$ and the associated cutoff size $s_o$
increase with disorder and we expect them to diverge as $o\rightarrow
0$. The kinetics of the approach towards ``infinite'' disorder depends on the specific distribution. For example, for the Weibull distribution we obtain the following limit of Eq. \eqref{eq:wei_p(T)}, valid for $p\in[0,1)$:
\beq
\lim_{k\rightarrow 0} T(p)=
\begin{cases}
  0 & \mbox{if } \; p < p_0 \\
  \lambda & \mbox{if } \; p = p_0 \\
  \infty & \mbox{if } \; p > p_0
\end{cases}
\label{eq:wei_T_limit}
\eeq
where $p_0=1-e^{-1}=0.63$. After an infinitesimal arrival time the
balls take the form of the percolation clusters obtained at
$p=p_0$. This is followed by an infinitely slow growth. At the limit
we have $B(T)=B(0)$ for $T>0$, where $B(0)$ is the percolation cluster
obtained at $p=p_0$.

For the Log-Normal distribution we have:
\beq
T(p)=e^\mu \exp{\left(\sqrt{2\sigma^2}\mbox{erf}^{-1}\left[2p-1\right]\right)}
\label{eq:logn_T(p)},
\eeq
where $\mbox{erf}^{-1}(x)$ is the inverse error function, and the
approach to the critical point is similar:
\beq
\lim_{\sigma\rightarrow \infty} T(p)=
\begin{cases}
  0 & \mbox{if } \; p < p_0 \\
  e^\mu & \mbox{if } \; p = p_0 \\
  \infty & \mbox{if } \; p > p_0
\end{cases}
\eeq
but now we have $p_0=1/2$, i.e. the critical ball is just the critical percolation cluster obtained at the critical probability.

Finally, for the Pareto distribution we have
\beq
T(p)=t_m\left(\frac{1}{1-p} \right)^{1/\alpha}
\label{eq:par_T(p)},
\eeq
and we obtain:
\beq
\lim_{\alpha\rightarrow 0} T(p)=\infty \quad \mbox{for } \; 0<p <1,
\eeq
which means that we must resort to increasingly large arrival times in order to observe the growth.

\subsection{Weak disorder regime $o\gg o^{\star}$}
\label{subsec:weak}

It is interesting to finish this work by considering the homogeneous
case in which all links have the same crossing time $\tau_0$. As
discussed in Sec. \ref{sec:model} and shown in Appendix
\ref{appendix:distrib}, this case corresponds to the limit
$o\rightarrow \infty$ of the link-time distributions, even though the
specific value of $\tau_0$ depends on the distribution function (see
Eq. \eqref{eq:homogeneous_case}): $\tau_0=\lambda, e^{\mu}, t_m$ for
Weibull, Log-Normal and Pareto respectively.

For the homogeneous case the chain model gives exactly $\ell_o(T)=T/\tau_0$.
Now we can use the following result, which applies to all
distributions:
\beq
\lim_{o \rightarrow \infty} {T (p)}=\tau_0 \quad \mbox{for } p<1,
\eeq
to obtain
\beq
\lim_ {o\rightarrow \infty} \ell_o(p) = 1 \quad \mbox{for } p<1.
\label{eq:homo_limit}
\eeq
It is important to note that this limit strictly holds for probabilities
smaller than 1. At the limit $p\rightarrow 1$ we obtain $\ell_o \rightarrow
\infty$.

This result confirms the assumption made in the previous section:
\beq
\lim_ {o\rightarrow \infty} \ell_o(p) -1 = \delta(1-p),
\label{eq:homo_limit_delta}
\eeq
and also shows that for a weak enough disorder we
cannot observe the percolation phase because it only
extends to a few sites.

As a numerical example, let us consider the distribution Wei$(1,3)$
where value $k=3$ is above the crossover value $k^{\star}\approx 1.79$. From
the chain model we obtain $\ell_3(p=0.99999)\approx 1.86$, which roughly
means that percolation effects up to a probability of $p=0.99999$ are
limited to a length scale of less than two lattice units. On the other
hand, the arrival time corresponding to that probability is
$T(p=0.99999)\approx2.26$. Now we can use the value of the mean time for
that distribution, $\tau\approx 0.89$ (close to $\tau_0=\lambda=1$), to
obtain $T/\tau\approx 2.53$, which provides a good estimate of
$\ell_3(p=0.99999)$. This result shows that we cannot observe
percolation effects even at disorders close to the crossover value
$o^{\star}$.


\section{Conclusions and further work}
\label{sec:conclusions}

In this work we have characterized the dynamics of first-passage
percolation (FPP) square lattices under extreme disorder, as opposed
to the weak-disorder regime, which is dominated by KPZ
universality. Several link-time distributions were considered which
allows for a continuous variation of the disorder strength through the
so-called order parameter $o$. A crossover value $o^{\star}$ (for
which $d_c=1$) was proposed as the crossover point between the two
regimes.

Our study has revealed that, given a certain level of disorder, there
exists a characteristic length scale $\xi_o$ below which the FPP model
behaves essentially as a bond-percolation lattice. Arrival times at
length scales smaller that $\xi_o$ follow a max principle which allow
us to establish a continuous mapping of FPP passage time $T$ into the
probability $p$ of a bond being open in bond-percolation. The basic
assumption is that the sum of link-times along the geodesic can be
approximated by their maximum value. As a result the mapping has the
form $p=F(T)$, where $F$ denotes the cumulative distribution function
for the link-times.

At length scales below $\xi_o$ the FPP model displays the same
criticality found in the second-order phase transition observed in
bond-percolation. The average value of the arrival time to different
sites becomes ill-defined and the geodesic between neighboring points
of the lattice can become arbitrarily large, instead of using a single
link as in the weak-disorder regime. Through a comprehensive scaling
analysis of the FPP balls we have been able to observe the critical
exponents characterizing the scaling of the percolation clusters near
the critical point, including their fractal structure, instead of the
reasonably rough circular balls within the KPZ universality class.

The dynamics of the FPP growth under strong disorder conditions is the
result of the interplay between this crossover length $\xi_o$ and the
correlation length $\xi$ characteristic of bond-percolation, both
evolving dynamically with passage time $T$. When $\xi \ll \xi_0$ the
growth of the FPP balls with passage time $T$ can be mapped into the
growth of the percolation clusters with increasing probability
$p=F(t)$, which resembles a sort of \emph{invasion percolation}. At a
certain passage time $T_o^{\star}$ (which is upper bounded by the
critical arrival time $T_c$), $\xi$ becomes of the same order of
$\xi_0$. Further growth leads to the failure of the maximality
assumption: the sum of the link-times along a geodesic may become
significantly different from the maximal value found along it. Thus
balls must start a rounding-off process that yields a crossover
towards KPZ-scaling. Therefore, for long enough distances we always
recover the KPZ regime.

The crossover length $\xi_o$ (as well as other related magnitudes
defined as a function of the order parameter) increases as the
disorder becomes stronger and seems to diverge when the order
parameter approaches zero, which we call the \emph{infinite disorder}
limit. The infinite disorder limit of these models presents very
intriguing features, which might be related to \emph{critical} or
\emph{supercritical} FPP cases, that we intend to ascertain in future
work. We have provided very preliminary evidences pointing to that
idea. Note, for example, that limit when $o\rightarrow 0$ of the
Weibull and Log-Normal distribution functions given in
Eqs. \eqref{eq:wei_f_F} and \eqref{eq:logn_f_F} are, for non-zero $t$,
$F(t)=1-e^{-1}=0.63$ and $F(t)=1/2$ respectively. They have the form
of the Bernoulli distribution (except for $t=0$) in which link-times
can be zero (open bonds) with probability $p_0$ and infinite (closed
bonds) with probability $1-p_0$, with $p_0$ given by the above
values. For the Weibull distribution we have $p_0>p_c$ and its
infinite-disorder limit might be related to critical FPP, whereas for
the Log-Normal we have $p_0=p_c$ and thus the supercritical
case. Since length $\xi_o$ determines the crossover length scale
between percolation and KPZ phases, and it diverges at the
infinite-disorder limit, certainly it would be very interesting to
study the criticality near this limit.

Another direction of future work is to elaborate more on the estimate
for $\xi_o$ derived from the chain model. Although the numerical
evidences presented here support its validity, there are still many
open questions that deserve additional work, including a theoretical
justification of Eq. \eqref{eq:CEF} and the generalization of the observed behaviors to other link-time distributions.


\begin{acknowledgments}
We thank R. Cuerno, E. Koroutcheva and E. Rodríguez-Fernández for very
useful discussions. Also, we acknowledge the Spanish government for
financial support through grant
PGC2018-094763-B-I00. D.V. acknowledges the Community of Madrid for the predoctoral contract PEJD-2018-PRE/IND-9095 funded by the Youth Employment Initiative (YEI). I.A.D. acknowledges the Community of Madrid for the research contract PEJ-2018-AI/IND-
10573 funded by the Youth Employment Initiative (YEI).
\end{acknowledgments}

\FloatBarrier


\appendix
\section{Link-time distributions}
\label{appendix:distrib}

The analysis performed in this work requires link-time
distributions with the following properties: (a) the link-times must
be always positive; (b) the range of disorder must be large, i.e. for
some range of the distribution parameters, the deviation must be
larger than the average value. We have chosen three distributions
fulfilling those requirements: Weibull, Log-Normal and Pareto. They
have been chosen for several reasons. First, their mathematical
expressions allow a simple analytic treatment. Also, we can easily
tune the strength of the disorder through a single parameter, the {\em
  shape parameter}. Besides, some of them (e.g. Weibull) generalizes a
number of other distributions. The mathematical expressions for the
probability density function $f(t)$ and the cumulative distribution
function $F(t)$ of those distributions are:

\begin{enumerate}
\item[(i)] The Weibull distribution, Wei$(\lambda,k)$:
  \beq
  \begin{aligned}
    f(t)&={k\over\lambda} \({t\over\lambda}\)^{k-1}\exp\(-(t/\lambda)^k\), \\
    F(t)&=1- \exp\(-(t/\lambda)^k\).
  \end{aligned}
  \label{eq:wei_f_F}
  \eeq
The results displayed in this work do not depend on
the scale parameter $\lambda$, so we have considered $\lambda=1$ in
all the numerical simulations. As we will see, it is the shape
parameter $k$ which completely determines the strength of the
disorder.

\item[(ii)] The Log-Normal distribution, LogN$(\mu,\sigma)$:
\beq
\begin{aligned}
f(t)&={1\over t\sqrt{2\pi \sigma^2}} \exp{\left[-{(\ln t-\mu)^2 \over 2\sigma^2}\right]},
\\
F(t)&=\frac{1}{2}+\frac{1}{2}\mbox{erf}\left[\frac{\ln t - \mu}{\sqrt{2\sigma^2}}\right].
\end{aligned}
\label{eq:logn_f_F}
\eeq
The above comment about the parameters of the Weibull distribution
also applies here to parameters $\mu$ and $\sigma$ respectively.

\item[(iii)] The Pareto distribution, Par$(t_m,\alpha)$:
\beq
\begin{aligned}
f(t)&=\begin{cases}
{\alpha t_m^\alpha\over t^{\alpha+1}} & \mbox{if } \; t\geq t_m \\
0 & \mbox{if } \; t<t_m
\end{cases}
\\
F(t)&=\begin{cases}
1- \left({t_m \over t}\right)^\alpha & \mbox{if } \; t\geq t_m \\
0 & \mbox{if } \; t<t_m
\end{cases}
\end{aligned}
\label{eq:par_f_F}
\eeq
with the above remark applying to scale $t_m$ and shape $\alpha$
parameters respectively.
\end{enumerate}

The homogeneous case discussed in Sec. \ref{sec:model} is
obtained at the following limits of the above distributions:
\begin{equation}
\label{eq:homogeneous_case}
\begin{aligned}
  \lim_{k\rightarrow \infty} \mbox{Wei}(\lambda,k) &= \delta(t-\lambda), \\
  \lim_{\sigma\rightarrow 0}
\mbox{LogN}(\mu,\sigma) &= \delta(t-e^{\mu}), \\
  \lim_{\alpha \rightarrow \infty}
\mbox{Par}(t_m,\alpha) &= \delta(t-t_m).
\end{aligned}
\end{equation}

A relevant parameter in our discussion is $d_c=(1/3)(\tau/s)^2$, where
$\tau$ is the mean value and $s$ the standard deviation of the
distributions (when they exist). For the Weibull distribution we find:
\beq
 d_c(k)={1\over 3}\frac{\Gamma^2 \left[1+{1\over k}\right]}{\Gamma
   \left[1+{2\over k}\right]-\Gamma^2 \left[1+{1\over k}\right]},
\label{eq:wei_dc}
\eeq
and the crossover value of $k$ for which $d_c=1$, denoted here as $k^{\star}$, is
\beq
  k^{\star}\equiv k(d_c=1)\approx 1.79.
\label{eq:wei_kc}
\eeq
For the Log-Normal distribution we have:
\beq
 d_c(\sigma)={1\over 3}\frac{1}{e^{\sigma^2}-1} \quad \mbox{with}
 \quad \sigma^{\star}\approx 0.54,
 \label{eq:logn_dc}
\eeq
and for Pareto:
\beq
 d_c(\alpha)={1\over 3}\alpha (\alpha-2)  \quad \mbox{with} \quad  \alpha^{\star}=3.
\label{eq:par_dc}
\eeq
Note that the last expression only holds for $\alpha>2$ since the standard deviation $s$ diverges when $\alpha\leq 2$ and the mean time $\tau$ diverges if $\alpha\leq 1$. For $\alpha\leq 2$ we shall assume that $d_c=0$.

In the three cases $d_c$ is a monotonic function of the shape
parameter: it decreases and approaches $0$ as the dispersion of the
distribution increases ($k\rightarrow 0$, $\sigma \rightarrow \infty$
and $\alpha\rightarrow 0$), and diverges as the distributions approach the
delta distribution ($k\rightarrow \infty$, $\sigma \rightarrow 0$ and
$\alpha\rightarrow \infty$).


\end{document}